\documentclass[a4paper,11pt]{article}
\pdfoutput=1

\usepackage{geometry}										
\usepackage[english]{babel}									
\usepackage[utf8]{inputenc}									
\usepackage[T1]{fontenc}									

\usepackage{lmodern}										

\usepackage{graphicx}
\usepackage[normalem]{ulem} 
\usepackage{xcolor}
\usepackage{siunitx}
\usepackage{placeins} 
\usepackage{bbold} 

\usepackage{amsmath}										
\usepackage{amstext}										
\usepackage{amsfonts}										
\usepackage{amssymb}										
\usepackage{amsthm,amscd}

\allowdisplaybreaks

\usepackage{authblk}										

\title{A spectral surrogate model for stochastic simulators computed from trajectory samples}
\author[1]{Nora L\"uthen}
\author[1]{Stefano Marelli}
\author[1]{Bruno Sudret}
\affil[1]{Chair of Risk, Safety, and Uncertainty Quantification, ETH Z\"urich,
	8093 Z\"urich, Switzerland}
\date{\today}

\usepackage[caption=false]{subfig}
\usepackage{caption}
\usepackage{natbib} 
\usepackage{hyperref}
\hypersetup{
	pdftitle = {},
	pdfauthor = {},
	pdfsubject = {},
	pdfkeywords = {},
	colorlinks = true,										
	linkcolor={red!90!black},
	citecolor={green!50!black},
	urlcolor={red!50!black}
}

\usepackage[capitalize]{cleveref}
\usepackage{enumitem} 
\setlist{nosep}
\usepackage{placeins} 
\usepackage{siunitx} 
\usepackage{bbold} 

\newcommand{\Nparams}{N}
\newcommand{\Nlatent}{R}
\newcommand{\Nparam}{N}
\newcommand{\Nactive}{P}
\newcommand{\Nkl}{M}

\newcommand{\alp}{{\ve \alpha}}
\newcommand{\ivsp}{{L^2_{f_{\ve X}}(\cd)}} 

\newcommand{\defineterm}[1]{\textit{#1}}

\newcommand{\norme}[2]{\left\lVert #1 \right\rVert_{#2}}

\newcommand{\ca}{{\mathcal A}}
\newcommand{\cb}{{\mathcal B}}
\newcommand{\cm}{{\mathcal M}}
\newcommand{\cd}{{\mathcal D}}
\newcommand{\cl}{{\mathcal L}}
\newcommand{\cf}{{\mathcal F}}
\newcommand{\cn}{{\mathcal N}}
\newcommand{\cg}{{\mathcal G}}
\newcommand{\co}{{\mathcal O}}
\newcommand{\ct}{{\mathcal T}}
\newcommand{\cu}{{\mathcal U}}
\newcommand{\cx}{{\mathcal X}}
\newcommand{\cz}{{\mathcal Z}}
\newcommand{\ve}[1]{\boldsymbol{#1}}			
\newcommand{\Rr}{{\mathbb R}}
\newcommand{\Nn}{{\mathbb N}}
\newcommand{\Prob}[1]{{\mathbb P}\left( #1 \right)}	
\newcommand{\enum}{ , \, \dots \,,}
\newcommand{\Esp}[1]{{\mathbb E}\left[ #1 \right]}
\newcommand{\Espe}[2]{{\mathbb E}_{#1}\left[#2\right]}
\newcommand{\Var}[1]{{\rm Var}\left[ #1 \right]}
\newcommand{\Vare}[2]{{\rm Var}_{#1}\left[#2\right]}
\newcommand{\Cov}[1]{{\rm Cov}\left[ #1 \right]}
\newcommand{\di}[1]{{\rm d}#1} 				

\newcommand{\covfct}{c}
\newcommand{\coeff}{a}
\newcommand{\eigv}{b}

\newtheoremstyle{bfnote}%
{}{}%
{}{}%
{\bfseries}{.}%
{ }%
{\thmname{#1}\thmnumber{ #2}\thmnote{ (#3)}}
\theoremstyle{bfnote}
\newtheorem{remark}{Remark}

\begin{document}
	\maketitle
	\begin{abstract}
		Stochastic simulators are non-deterministic computer models which provide a different response each time they are run, even when the input parameters are held at fixed values.
		They arise when additional sources of uncertainty are affecting the computer model, which are not explicitly modeled as input parameters.
		The uncertainty analysis of stochastic simulators requires their repeated evaluation for different values of the input variables, as well as for different realizations of the underlying latent stochasticity.
		The computational cost of such analyses can be considerable, which motivates the construction of surrogate models that can approximate the original model and its stochastic response, but can be evaluated at much lower cost.
		
		We propose a surrogate model for stochastic simulators based on spectral expansions. 
		Considering a certain class of stochastic simulators that can be repeatedly evaluated for the same underlying random event, we view the simulator as a random field indexed by the input parameter space. 
		For a fixed realization of the latent stochasticity, the response of the simulator is a deterministic function, called trajectory.
		Based on samples from several such trajectories, we approximate the latter by sparse polynomial chaos expansion and
		compute analytically an extended Karhunen-Lo\`eve expansion (KLE) to reduce its dimensionality.
		The uncorrelated but dependent random variables of the KLE are modeled by advanced statistical techniques such as parametric inference, vine copula modeling, and kernel density estimation. 
		The resulting surrogate model approximates the marginals and the covariance function, and allows to obtain new realizations at low computational cost.
		We observe that in our numerical examples, the first mode of the KLE is by far the most important, and investigate this phenomenon and its implications. 
	\end{abstract}

\section{Introduction}
Nowadays, computer simulations are an essential ingredient of the research and development workflow in virtually all fields of science and engineering.
Typically, not all parameters and conditions needed for the simulations are known exactly, and this uncertainty affects the output of the simulations. This is the main focus of the field of uncertainty quantification \citep{SmithUQBook2014}. 

Most computer simulations can be classified as \defineterm{deterministic simulators}: repeatedly evaluating the model $\cm$ for the same set of input parameters $\ve x$ always yields the same deterministic response $y = \cm(\ve x) \in \Rr$.%
\footnote{We consider here only real-valued simulators. The extension to low-dimensional vector-valued simulators is straightforward. For the extension to high-dimensional vector-valued or function-valued simulators, see e.g.\ \citet{NagelEAWAG2020,Perrin2021}.}
To perform uncertainty quantification, the uncertainty on the input (parameters and conditions) is usually represented probabilistically, and we follow this approach in this paper.
Propagating the input uncertainty through the deterministic simulator, the overall response of the simulation becomes a random quantity.

However, not all computer simulations can be classified as deterministic simulators. Some models contain intrinsic stochasticity that cannot be modeled as input parameter, e.g., epidemiological models where each transmission or recovery is a random event, governed by the respective rate of occurrence. Other models depend on uncontrollable environmental variables such as wind fields or earthquakes, for which it can be infeasible or undesirable to explicitly model their uncertainty.
In these cases, it is more convenient to use the notion of a \defineterm{stochastic simulator}: only some of the uncertainty is explicitly modeled as random input variables, and there is some residual randomness affecting the computational model that causes the model response $\cm(\ve x)$ for a fixed set of input parameters $\ve x$ to still be a random variable: $Y_{\ve x} = \cm(\ve x)$. In other words, evaluating the computer model several times with the same input parameters $\ve x$ will result in different realizations $y$ of the random variable $Y_{\ve x}$.
Of course, since there is no true randomness in a computer, every computer simulation  can be made deterministic by fixing the random seed. However, the seed is in general not a useful parametrization of uncertainty.

Uncertainty quantification methods typically require many runs of the computational model, which can become costly or even infeasible for expensive engineering simulators. 
To save computational resources, the model is often replaced with a cheaper \defineterm{surrogate model} (or \defineterm{metamodel}), which provides a reasonably good approximation to the original model. The surrogate model is computed from a small number of model evaluations and can subsequently be evaluated many times with negligible computational cost.
Surrogate models often treat the model as a \defineterm{black box}, i.e., they do not use any specific knowledge about the model and rely only on the available input-output data samples (and sometimes on the characteristics of the input parameter space). 
Popular surrogate models for deterministic simulators include 
polynomial chaos expansions \citep{Ghanembook1991,Xiu2002}, 
Kriging \citep{Sacks1989, Rasmussen2006}, 
radial basis functions \citep{Buhmann2000},
and support vector regression \citep{Vapnik:1995, Smola2004}.

Since the response of stochastic simulators is a random variable for every set of input parameters, even more runs might be required to analyze their uncertainty, making surrogate models all the more relevant in this case. 
Research on surrogating stochastic simulators is comparatively recent. 
Most available methods focus on the marginal response distribution $\Prob{Y | \ve X = \ve x}$ for $\ve x \in \cd$ 
and emulate the conditional density itself or certain statistics of it.
Early contributions aimed at characterizing the variation of the first two moments of the output response over the input domain using joint Gaussian process models \citep{Iooss2009,Marrel2012}.
Another class of methods aims at directly modeling the variation of the marginal output probability density function (pdf) of the random variable $Y_{\ve x}$ over the input domain. 
Assuming that the true marginal response pdf at a number of input locations is known, \citet{Moutoussamy2015} represent the marginal pdf of a new input point as a linear combination of training examples (i.e., kernel regression) or of specifically constructed basis functions. However, the true marginal pdf is rarely known or its generation might require a lot of samples.
For a finite number of stochastic simulator evaluations over the input domain (with or without replications), \citet{ZhuIJUQ2020, ZhuSIAM2021} model the variation of the marginal output pdf over the input domain using the so-called \defineterm{generalized lambda model}, a parametric distribution family that is able to approximate many classical families.
In fact, stochastic simulators are akin to real-world scientific experiments, which are usually stochastic due to unavoidable measurement error and environmental noise. Therefore, standard statistical methods like quantile regression \citep{Torossian2020} and kernel conditional density estimation \citep{Hall2004} can also be used to emulate the marginal distribution of the response of a stochastic simulator.
Furthermore, \citet{ZhuStoPCE2022} developed an approach that emulates the stochastic simulator response in distribution, inspired by the weak PCE methodology based on maximum likelihood estimation \citep{XiuBook2010}.

A related method from machine learning are Bayesian neural networks, whose weights are modeled as independent Gaussian random variables \citep{MacKay1992b,Goan2020}. Bayesian methods such as Markov Chain Monte Carlo or variational inference are used to determine the parameters of the weight densities from the given data.
Furthermore, generative models like variational autoencoders \citep{Kingma2014} and generative adversarial networks \citep{Goodfellow2014} can be seen as surrogate models in distribution, learning a conditional target density from data.

All the methods cited above aim at emulating only the univariate probability density functions of the response random variables of the stochastic simulator. 
However, they do not take into account the correlation and higher-order information between the stochastic simulator responses at different points in the input domain. 
This close relation between the responses at different input locations can be best illustrated by fixing the stochasticity of the simulator (e.g., by fixing the random seed)%
\footnote{Note that this does not require this randomness to be modeled. In practice, fixing the seed might not be possible for all computational models, as it depends on their implementation.}%
: in this case, the stochastic simulator response over the input domain becomes a deterministic function, which we call a \defineterm{trajectory}.
In other words, the stochastic simulator can be seen as a random field, i.e., as 
a collection of random functions.

Surrogating a stochastic simulator based on few model evaluations becomes therefore the task of inferring a random field from discrete samples (often called ``limited data'').
Popular methods for modeling random fields include orthogonal series expansions, such as spectral representation \citep{Shinozuka1991, Grigoriu1993} or Karhunen-Lo\`eve expansion (KLE) \citep{Loeve1978, Karhunen1946,Zhang94,Ramsay2005,Grigoriu2006}, and translation processes, which are mappings of Gaussian processes \citep{Yamazaki1988, Grigoriu1998, Sakamoto2002,Shields2011}. 
To our knowledge, the only publication in the specific context of stochastic simulators which takes the random field point of view and aims at emulating trajectories (including the higher-order relations between responses at different input locations) is by \citet{AzziIJUQ2019}, who construct a metamodel using Karhunen-Lo\`eve expansion together with the deterministic methods PCE and Kriging.

The goal of our paper is to develop a surrogate model that is able to emulate the trajectories of a stochastic simulator, and allows insight into the dependence between the simulator responses at different input locations.
Our method of choice in this paper is Karhunen-Lo\`eve expansion, one of the most popular methods for random field inference from limited data.
The main challenges in constructing a trajectory-based surrogate for a stochastic simulator (a \defineterm{stochastic emulator}) are explained in more detail in the following:
\begin{enumerate}
	\item Accuracy and efficiency: the surrogate should be accurate while needing as few model evaluations as possible.
	\label{chall:effic}
	\item Continuous surrogate from discrete data:%
	\label{chall:interp}
	the surrogate should emulate the response over the whole (continuous) input domain, while the available data consists of trajectories sampled at a few points throughout the input domain (i.e., discrete samples). 
	\item The stochastic simulator is in general a non-Gaussian random field. This introduces additional complexity into the Karhunen-Lo\`eve model. 
	\label{chall:non-gauss}
\end{enumerate}

We are addressing each of these challenges by introducing a novel approach that combines several state-of-the-art methodologies.
We use Karhunen-Lo\`eve expansion in conjunction with sparse PCE \citep{BlatmanJCP2011, LuethenSIAMJUQ2021}, which is a powerful and sample-efficient surrogate modeling method for deterministic simulators, to address Challenge~\ref{chall:effic}. This circumvents the otherwise high computational cost of solving the integral eigenvalue problem of KLE \citep{Schwab2006, Betz2014} by reducing the integral eigenvalue problem to finite-dimensional discrete principal component analysis (PCA) in the truncated space of PCE coefficients.
The joint distribution of the resulting sample of dependent random KLE coefficients (Challenge~\ref{chall:non-gauss}) is identified using statistical inference within the marginal-copula framework \citep{TorrePEM2019}.
The procedure results in an analytical formula for the stochastic emulator that can be used for computing marginals and correlations, as well as for generating new trajectories that resemble trajectories of the original stochastic simulator.

In our approach, the extension from discrete data to the continuous model (Challenge~\ref{chall:interp}) is achieved by approximating the sampled trajectories by sparse regression-based PCE. 
A similar approach has been used by \citet{Navarro2017} in the context of stochastic differential equations with the goal of sensitivity analysis, using non-intrusive pseudospectral projection to compute the PCE coefficients.
The representation by sparse PCE can be seen as a variant of \defineterm{orthogonal series expansion (OSE)} \citep{Zhang94}, which expands a second-order random process in terms of an orthogonal basis of the associated Hilbert space. 

Note that when random fields are approximated based on a set of samples, it is most often assumed that the latter are collected on a discrete mesh in the index set, whereas this is not a requirement for our method.
In such mesh-based approximations to random fields, PCE is often used for modeling the random variables arising in dimension-reduced expansions \citep{Desceliers2006a, Doostan2007, Das2009, Raisee2015, Abraham2018, Dai2019}.
This is distinct from our approach, as we use PCE to approximate the trajectories in the input space.
Our approach yields an emulator for the whole input space (including unseen locations), while existing approaches are mostly focused on building an emulator on the discrete mesh where the samples were collected. 

KLE represents a random field using an optimal orthogonal basis of the index space,
resulting in an expansion in terms of deterministic functions weighted by random coefficients.
These random coefficients are by construction uncorrelated, but unless the random field is a Gaussian random field, they are in general statistically dependent.
Inferring the joint distribution of dependent random variables from samples is challenging but necessary for approximating a general non-Gaussian random field by KLE.
To address this challenge of inference, several approaches have been proposed. \citet{Grigoriu2010b} suggests two methods to infer the joint distribution of the random coefficients of a series expansion model, of which one amounts to kernel density estimation, and the other to the fitting of a discrete joint distribution.
\citet{Poirion2013, Poirion2014} use KLE for modeling seismic ground motion time series, and model the random KLE coefficients by 1D sample CDFs assuming at most pairwise dependence \citep{Poirion2013}, or by kernel density estimation \citep{Poirion2014}. 
In the present paper, we investigate the use of kernel density estimation and inference of parametric joint distributions based on marginals and vine copulas.

This paper is organized as follows: in \cref{sec:theory} we recall the relevant theory and definitions.
In \cref{sec:our_approach} we present our new stochastic emulator. 
The proposed method is then applied in \cref{sec:numerical_experiments}, where we assess its performance on several examples of varying complexity.
Here we observe that the KLE is often significantly dominated by its first mode, a phenomenon that we investigate in \cref{sec:numbermodes}.
Finally, we draw conclusions and give an outlook on possible further developments in \cref{sec:conclusion}.

\section{Theoretical foundation}
\label{sec:theory}
We provide a brief summary of the relevant theory and concepts needed to construct our proposed stochastic emulator for stochastic simulators: 
random fields, polynomial chaos expansions, Karhunen-Lo\`eve expansion, and inference of joint probability distributions.

\subsection{Stochastic simulators as random fields}
\label{sec:stochsim}
\label{sec:random_fields}

\begin{figure}[htb]
	\subfloat[Random variable $Y_{\ve x} = \cm(\ve x, \cdot)$]{\includegraphics[width=.47\textwidth]{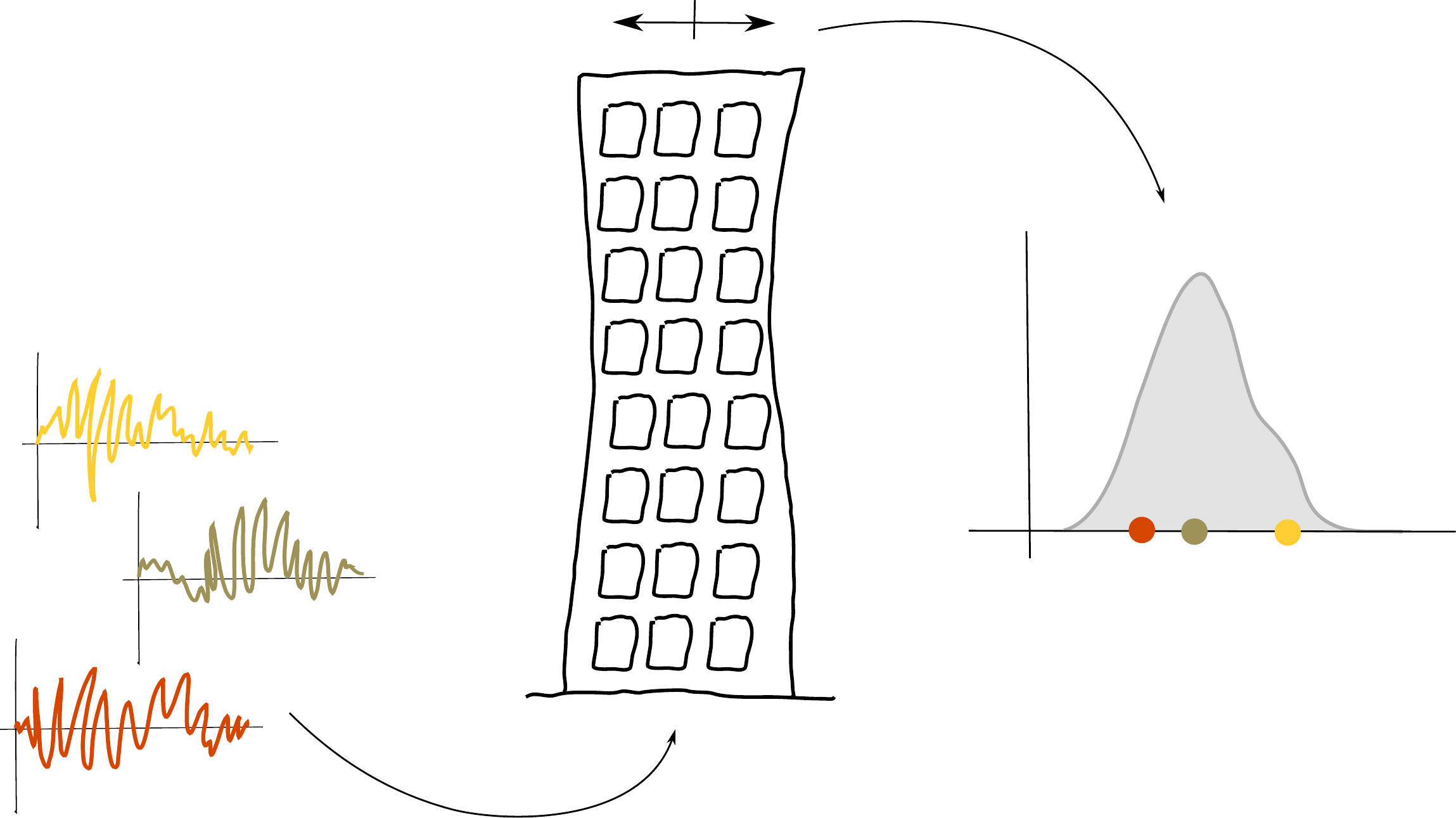}}
	\hfill
	\subfloat[Trajectory $\cm(\cdot, \omega): \cd \to \Rr$]{\includegraphics[width=.49\textwidth]{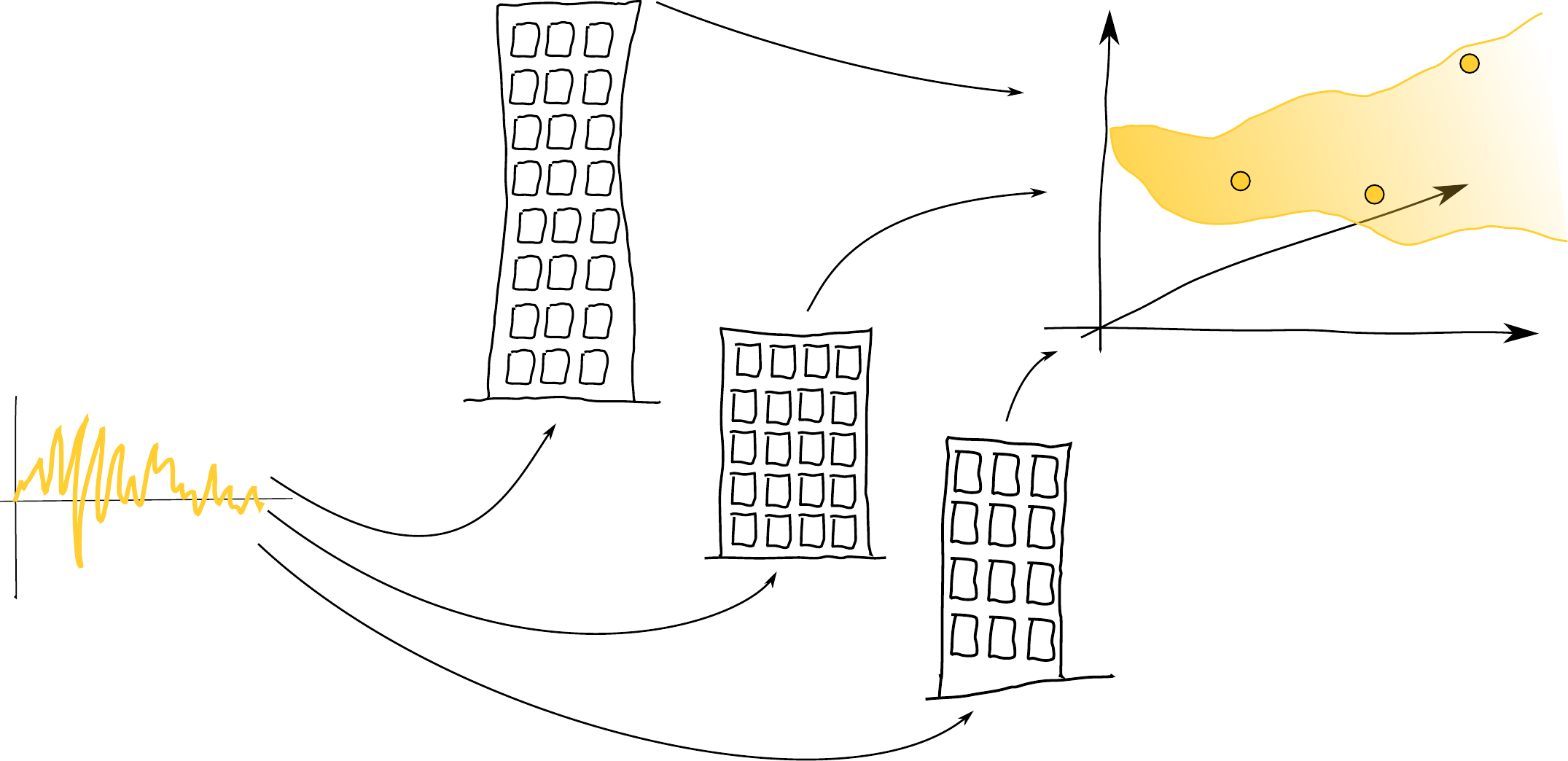}}
	\caption{Visual representation of a stochastic simulator when either the input parameters $\ve x$ or the random event $\omega$ are held fixed, resulting in a random variable (left) or a deterministic function (right). 
		The computational model is a high-rise building parametrized by several properties $\ve x$ (visualized in the sketch by the shape of the building) subject to random earthquake events $\omega$ (visualized by 1D time series in different colors), whose output $\cm(\ve x, \omega)$ 
		is a real number (e.g., the maximal displacement at the top floor). 
	}
	\label{fig:visualize_stochastic_simulator}	
\end{figure}

Let $\ve X$ be a random vector with values in $\cd \subset \Rr^d$, with finite variance and joint probability density function (pdf) $f_{\ve X}$.
Denote by $\omega \subset \Omega$ an abstract random event in a probability space $(\Omega, \cf, P)$.
A \defineterm{stochastic simulator} is a mapping
\begin{align}
\cm: \cd \times \Omega &\to \Rr, \\ 
(\ve x, \omega) &\mapsto \cm(\ve x, \omega).
\end{align}
Fixing $\ve x \in \cd$, the quantity $Y_{\ve x} = \cm(\ve x, \cdot): \Omega \to \Rr$ is a random variable.
Fixing $\omega \in \Omega$, $\cm(\cdot, \omega): \cd \to \Rr$ is a function in the input parameters, which we call \defineterm{trajectory} or \defineterm{realization} of the stochastic simulator (see also \cref{fig:visualize_stochastic_simulator}).
We assume that $Y_{\ve x}$ has finite variance for all $\ve x$, 
and that $\cm(\cdot, \omega) \in \ivsp$ for all $\omega \in \Omega$. 

These definitions imply that a stochastic simulator $\cm$ can be seen as a \defineterm{random field} (also: \defineterm{stochastic process} or \defineterm{random process}) $\{Y_{\ve x}\}_{\ve  x \in \cd}$ with \defineterm{index set} $\cd$, i.e., as a family of random variables $\{Y_{\ve x}\}$ indexed by $\ve x \in \cd$.
In the following, we provide a brief reminder of a few random field basics. For more details, see e.g.\ \citet{Grigoriu2002book}.

To fully characterize a general random field, one needs to specify the collection of all its finite-dimensional distributions
\begin{equation}
F_{Y_{\ve x_1} \enum Y_{\ve x_n}}(y_1 \enum y_n) = \Prob{Y_{\ve x_1} \leq y_1 \wedge \ldots \wedge Y_{\ve x_n} \leq y_n}
\end{equation}
for \textit{all} $n\geq 1$ and \textit{any} $\ve x_1 \enum \ve x_n \in \cd$.
Extending the concept of moments of random variables to random fields, 
the deterministic \defineterm{mean function} of the random field is given by $\mu(\ve x) = \Esp{Y_{\ve x}}$.
If $\mu(\ve x) = 0$, the random field is called \defineterm{centered}.
The \defineterm{(auto-)covariance function} is defined by
\begin{equation}
\covfct(\ve x, \ve x') = \Esp{(Y_{\ve x} - \mu(\ve x))(Y_{\ve x'} - \mu(\ve x'))}.
\label{eq:covariance_function}
\end{equation}

In general, a random field is not uniquely defined by its mean and covariance function.
The only exception is the family of \defineterm{Gaussian processes}, for which all finite-dimensional joint distributions are multivariate Gaussian distributions. 
For Gaussian processes, conditional distributions are again multivariate Gaussians, which lies at the foundation of the popular surrogate modeling technique Kriging/Gaussian process modeling. 
While Gaussian random fields are computationally convenient, random fields encountered in real-world problems (and in particular, stochastic simulators) are often non-Gaussian. One obvious argument is that Gaussian variables are unbounded while physical quantities are almost always bounded \citep{Grigoriu2002book}.

A special feature of a stochastic simulator $\cm$, as opposed to classical random fields, is that its index set is not an interval or a hypercube, but a general domain $\cd \in \Rr^d$ with weight function $f_{\ve X}$. We will use this property to build an accurate surrogate model for $\cm$ respecting the probability density $f_{\ve X}$ of the input space.

\subsection{Polynomial chaos expansion}
\label{sec:PCE}
Polynomial chaos expansion (PCE) is a technique for modeling random variables using a basis of polynomials that are orthonormal w.r.t.\ a given probability density function \citep{Ghanembook1991, Xiu2002}.
In our algorithm (\cref{sec:our_approach}), we will use PCE to approximate trajectories $\cm(\cdot, \omega)$ of the stochastic simulator, which can be seen as random variables $\cm(\ve X, \omega)$ with their randomness induced by the uncertainty in the input $\ve X$.

Consider a random vector $\ve X$ with values in $\cd \subset \Rr^d$ and independent components. Let $f_{\ve X}(\ve x) = \prod_{i=1}^d f_{X_i}(x_i)$ be its probability density function (pdf) and assume that $\ve X$ has finite variance. 
Let $\ivsp$ be the space of real-valued function that are square-integrable under $f_{\ve X}$, i.e., $\ivsp = \left\{g:\cd \to \Rr \ \big| \ \Vare{\ve X}{g(\ve X)} < +\infty \right\}$.
Under certain assumptions on the random vector $\ve X$ \citep{Xiu2002, Ernst2012}, there exists an orthonormal polynomial basis $\{\psi_\alp \ | \ \alp \in \Nn^d\}$ of $\ivsp$,
where each element is the product of univariate polynomials characterized by the multi-index $\alp$.

Let $\cm \in \ivsp$ be a (computational) model. Its output $Y = \cm(\ve X)$ is a random variable, which can be represented in terms of the orthonormal polynomial basis as
$	\cm(\ve X) = \sum_{\alp \in \Nn^d} \coeff_\alp \psi_\alp(\ve X)$
with
$\coeff_\alp = \Espe{\ve X}{\cm(\ve X) \psi_\alp(\ve X)}$. 
This representation is called \defineterm{polynomial chaos expansion}. In practice, a truncated expansion is computed, 
\begin{equation}
\cm(\ve X) \approx \cm^\text{PCE}(\ve X) = \sum_{\alp \in \ca} \coeff_\alp \psi_\alp(\ve X),
\end{equation}
where $\ca \subset \Nn^d$ is a finite subset of the full basis.
The accuracy of a truncated PCE depends on three ingredients: the choice of $\ca$, the method used for computing the coefficients $\ve \coeff = (\coeff_\alp)_{\alp \in \ca}$, and the choice of points $\cx \subset \cd$ used in the coefficient computation method. An extensive overview of the state-of-the-art methods to determine these is given in \citet{LuethenSIAMJUQ2021,LuethenIJUQ2022}.

\subsection{Karhunen-Lo\`eve expansion}
\label{sec:KLE}
Karhunen-Lo\`eve expansion (KLE) is a well-established spectral expansion technique through which a random field is represented in terms of an optimal orthogonal basis for the index space, weighted by random coefficients
\citep{Karhunen1946, Loeve1978}.
KLE transforms the random field, which is an uncountably infinite but correlated family of random variables $\{\cm_{\ve x}\}_{\ve x \in \cd}$,
into a countably infinite but uncorrelated family of different random variables $\{\xi_i\}_{i = 1,2,\ldots}$. 
Furthermore, the random variables $\xi_i$ are typically of decreasing importance.
KLE is therefore well suited and often used for discretization and modeling efforts for random fields.

To make these notions more precise, 
let $\{\cm_{\ve x}(\omega)\}_{\ve x \in \cd}$ be a random field. Denote by $\mu(\ve x) = \Esp{\cm_{\ve x}}$ its mean function, and by $\covfct(\ve x, \ve x') = \Cov{\cm_{\ve x}, \cm_{\ve x'}}$ its covariance function. 
Let $\cd$ be closed and bounded. Let $\covfct$ be continuous on $\cd \times \cd$ and assume that $\cm_{\ve x}$ has finite variance for all $\ve x \in \cd$. 
Then the \defineterm{Karhunen-Lo\`eve expansion} of the random field $\cm_{\ve x}$ is given by
\begin{equation}
\cm_{\ve x}(\omega) = \mu(\ve x) + \sum_{k=1}^\infty\sqrt{\lambda_k}\xi_k(\omega) \phi_k(\ve x)
\label{eq:KLE}
\end{equation}
where $(\phi_k)_{k=1,2,\ldots}$ is an orthonormal basis of $L^2(\cd)$, $\lambda_1 \geq \lambda_2 \geq \ldots \geq 0$ is a non-increasing sequence of non-negative real numbers, and $\{\xi_k\}_{k=1,2,\ldots}$ is a countable family of  zero mean, unit variance, uncorrelated random variables.

Here, $(\lambda_k, \phi_k)$ are solutions to the integral eigenvalue problem
\begin{equation}
\int_{\cd} \covfct(\ve x, \ve x') \phi_k(\ve x') \di{\ve x'} = \lambda_k \phi_k(\ve x), 
\label{eq:integral_EV_problem}
\end{equation}
and
$\xi_k$ is the result of the projection of $\cm$ onto the spatial basis
\begin{equation}
\xi_k(\omega) = \frac{1}{\sqrt{\lambda_k}} \int_{\cd} \cm(\ve x, \omega) \phi_k(\ve x) \di{\ve x}.
\label{eq:KLE_projection}
\end{equation}
From \cref{eq:KLE} and the properties of $\phi_k$ and $\xi_k$ it follows immediately that the covariance function can be expressed as
\begin{equation}
\covfct(\ve x, \ve x') = \sum_{k = 1}^{\infty} \lambda_k \phi_k(\ve x) \phi_k(\ve x')
\label{eq:covfct_series}
\end{equation}
(Mercer's theorem).
Note that the KLE random variables $\{\xi_k\}$ (herein \defineterm{KL-RV}) do not enter this expression.

KLE is especially well-suited to Gaussian random fields, since in this case the random variables $\xi_k$ are standard Gaussian and independent.
However, \cref{eq:KLE} holds for all random fields fulfilling the assumptions, not only for Gaussian random fields. The non-Gaussianity is modeled by the (possibly complex) joint distribution $f_{\ve \xi}$ of the KL-RV.

\cref{eq:KLE,eq:integral_EV_problem,eq:KLE_projection} are formulated in terms of $L^2(\cd)$, but they can be generalized:
let $\ve X$ be a random variable with values in $\cd \subset \Rr^d$, density $f_{\ve X}$, and finite variance. Then KLE can be generalized to the space $\ivsp$ instead of $L^2(\cd)$. In that case, the index set $\cd$ does not have to be bounded, since the volume of $\cd$ under measure $f_{\ve X}\di{\ve x}$ is finite. This is called \defineterm{extended KLE} \citep{Iemma2006}.
This property is crucial for our proposed stochastic emulator, which we will introduce in \cref{sec:our_approach}.

In practice, the infinite expansion in \cref{eq:KLE} must be truncated. 
From the orthonormality of $\{ \phi_k\}$ it follows from \cref{eq:covfct_series} that the  
variance of the random field is equal to $\sum_{k=1}^\infty \lambda_k$. 
The sequence $\lambda_1 \geq \lambda_2 \geq \cdots \geq 0$ is non-increasing, and typically (depending on the correlation length of the random field) this sequence decays rather quickly to zero. Loosely speaking, the higher the correlation between different locations in the index set, the fewer spatial basis functions are needed to approximate the trajectories, and the faster the decay of the eigenvalues.
Knowing this, the KLE can be truncated at the $\Nkl$-th term with $\Nkl$ chosen so that the fraction of \defineterm{explained variance} is sufficiently large:
\begin{equation}
\frac{\sum_{k = 1}^\Nkl \lambda_k}{\sum_{k = 1}^\infty \lambda_k} > 1 - \epsilon
\label{eq:KLEthreshold}
\end{equation}
for a small threshold parameter $\epsilon > 0$ (e.g., $\epsilon = 0.001$).

KLE is closely related to function principal component analysis (fPCA) \citep{Besse1986, Ramsay2005}. To compute a solution to the integral eigenvalue problem in \cref{eq:integral_EV_problem}, there are several possibilities \citep[Section~8.4]{Ramsay2005}: the integrals can be approximated numerically; the eigenproblem can be discretized on a number of representative grid points in $\cd$ (this is the approach chosen by the majority of modelers, including \citet{AzziIJUQ2019}); or the eigenproblem can be written in terms of a suitable (truncated) spatial basis, which transforms the problem into a (finite-dimensional) discrete eigenvalue problem.
The third approach is related to orthogonal series expansion (OSE) \citep{Zhang94}.
It is used by \citet{Poirion2014}, who derive the explicit discrete problem for a basis consisting of interpolation functions, building on results by \citet{Besse1986} and \citet{Besse1991}.
We use this approach together with the orthogonal basis provided by polynomial chaos expansion (\cref{sec:our_approach}). Detailed calculations are provided in \cref{app:derivations_KLE_PCE}.

\subsection{Inference of the joint distribution of the Karhunen-Lo\`eve random variables}
\label{sec:inference_distributions}
Characterizing the dependent (but uncorrelated) Karhunen-Lo\`eve random variables (KL-RV) $\xi_k, k = 1 \enum \Nkl$ correctly is important for the accurate modeling of a general non-Gaussian stochastic process \citep{Grigoriu2010b}.
However, inferring the joint distribution of a random vector is a challenging task. 
The main challenge is the scarcity of data: the higher the dimensionality, the more samples are needed to be able to correctly infer the dependence structure of the data.
We need to construct a suitable parametric or non-parametric model to accurately describe the joint distribution.
In the following, we introduce the marginal-copula framework, which is a powerful tool to represent and infer complex dependence structures between random variables \citep{Nelsen2006,TorreJCP2019}.

Let $\ve Z$ be any $M$-dimensional random vector with multivariate cumulative distribution function (CDF) $F_{\ve Z}$ and marginal distributions $F_{Z_i}$. 
The so-called \defineterm{Sklar's theorem} states that $F_{\ve Z}$ can be written as
\begin{equation}
F_{\ve Z}(z_1 \enum z_M) = C\left(F_{Z_1}(z_1) \enum F_{Z_M}(z_M)\right),
\label{eq:CDF_copula_marginals}
\end{equation}
where the function $C:[0,1]^d \to \Rr$ is called \defineterm{copula} \citep{Sklar1959, Nelsen2006}. $C$ is a CDF with uniform marginals, which defines the dependence structure of the random vector $\ve Z$. $C$ is unique if all marginals $F_{Z_i}$ are continuous, and it holds that
\begin{equation}
C(u_1 \enum u_M) = F_{\ve Z}\left(F_{Z_1}^{-1}(u_1) \enum F_{Z_M}^{-1}(u_M)\right).
\label{eq:copula_CDF_marginals}
\end{equation}

Let an i.i.d.\ sample $\cz = \{\ve z^{(1)} \enum \ve z^{(N)}\}$ of the random vector $\ve Z$ be given. 
The goal is to infer the joint distribution $F_{\ve Z}$ from this sample. 
For this, the copula representation of \cref{eq:CDF_copula_marginals} is convenient, since it allows inferring the marginals and the dependence structure of the data separately, as briefly explained in the following.

To infer the marginal distributions, we consider two options. The first is parametric inference, where we choose from a set of parametric probability distributions with zero mean and unit standard deviation (see \cref{tab:marginals}) the distribution with the smallest Akaike information criterion (AIC).
If a distribution family has more than two parameters, its remaining parameters are chosen by maximum likelihood.
We utilize the 
uncertainty quantification software UQLab \citep{MarelliUQLab2014,UQdoc_14_114} with a modification prescribing the desired moments.

\begin{table}[htb]
	\centering
	\small
	\caption{Considered marginal families with zero mean and unit standard deviation.  
		The last column lists the remaining degrees of freedom $k$ after fixing the first two moments. The Akaike information criterion is then given as $\text{AIC} = 2k - 2\log{\cl}$, where $\cl$ is the likelihood.}
	\label{tab:marginals}
	\begin{tabular}{l p{.4\textwidth} l}
		\hline
		Type & Parameter  & $k$ \\
		\hline
		Uniform $\cu([a,b])$ & $a = -\sqrt{3}, b = \sqrt{3}$ & 0 \\
		Gaussian $\cn(\mu, \sigma)$ & $\mu = 0,\sigma = 1$ & 0 \\
		Gumbel (for maxima) $\cg(\mu, \beta)$ & $\mu \approx -0.4501, \beta \approx 0.7797$ & 0 \\
		Gumbel (for minima) $\cg_\text{min}(\mu, \beta)$ & $ \mu \approx 0.4501, \beta \approx 0.7797$ & 0 \\
		Logistic $P(\mu, s)$ & $ \mu = 0, s \approx 0.5513$ & 0 \\
		Laplace $\cl(\mu, b)$ & $ \mu = 1, b = \frac{1}{\sqrt{2}}$ & 0 \\
		Beta $\cb(a,b,r,s)$ & $a,b$ chosen according to data bounds \newline $r = \frac{a(ab+1)}{b-a}, s = \frac{b(ab+1)}{a-b}$ & 2 \\
		\hline
	\end{tabular}
\end{table}

A second popular method to represent marginal behavior non-parametrically is \defineterm{kernel density estimation} (KDE) \citep{Wand1995,Simonoff1996}, which has also been proposed for estimating the distribution of KL-RV \citep{Grigoriu2010b, Poirion2014}. Here the distribution is modeled as a Gaussian mixture, where the Gaussian density functions are centered in the data points and share the same standard deviation, called \defineterm{bandwidth} in the case of 1D KDE. 
We adopt a bandwidth estimation method optimal for data with Gaussian distribution \citep{Bowman1997}.

To characterize the dependence structure, we use a copula.
While any multivariate CDF with uniform marginals $\cu([0,1])$ constitutes a copula, there are a number of well-known parametric families (see, e.g., \citet{Nelsen2006, Joe2014}).
Besides the independence copula and the families derived from multivariate elliptical distributions, most of these parametric families are pair copulas, i.e., they couple only two variables.
Constructing meaningful parametric copulas for more than two variables (other than elliptical copulas) is in general difficult \citep{Nelsen2006}.

A solution is to decompose the $M$-variate copula into a product of conditional pair copulas, which is known as vine copula construction \citep{Bedford2002}. This is always possible as a consequence of the chain rule of probability. 
In general, a vine copula is the product of $\frac{M(M-1)}{2}$ pair copulas.%
\footnote{There are $M-1$ unconditional pair copulas; $M-2$ pair copulas conditioned on 1 other variable; $M-3$ conditioned on 2 other variables; and so on, until there is 1 pair copula conditioned on all except 2 variables.}
The factorization into pair copulas is not unique but depends on the ordering and grouping of variables. Two classes of vine copulas, differing in the order in which the variables are grouped into pairs, are the drawable vine (D-vine) \citep{Kurowicka2005} and the canonical vine (C-vine) \citep{Aas2009}. 
For a more detailed description of the vine copula construction, we refer to \citet{Aas2009} and \citet{TorrePEM2019}.

To infer a copula from data, we first map the multivariate data to $[0,1]^d$ by applying element-wise the inferred marginal CDFs (see \cref{eq:CDF_copula_marginals}). Then we infer the dependence structure by using Kendall's tau to determine the groupings of variables as well as their order in the vine copula \citep{Aas2009, TorrePEM2019}. For each pair copula, the parameters are identified by maximum likelihood. Finally, the best-fitting copula is chosen using AIC.
This approach is implemented in the statistical inference module of UQLab \citep{UQdoc_14_114}.
The list of available copula families can be found in \citet[Section~1.4]{UQdoc_14_102}.

\section{Surrogating a stochastic simulator from a set of samples}
\label{sec:our_approach}

We are now ready to describe the construction of our spectral surrogate model for a stochastic simulator.
Assume that discrete samples of the stochastic simulator $\cm$ are available in the following form: 
\begin{equation}
\ct_r = \left\{ \left( \ve x^{(r, i)} , \cm\big(\ve x^{(r, i)} , \omega^{(r)}\big)\right) : i = 1 \enum \Nparam_r \right\}, \quad r = 1 \enum \Nlatent
\label{eq:discrete_trajectories}
\end{equation}
i.e., in the form of discrete evaluations of the stochastic simulator on $\Nlatent$ trajectories,
where for every $r$, $\{\ve x^{(r, i)}: i = 1 \enum \Nparam_r\}$ is an i.i.d.\ sample from the input distribution $f_{\ve X}$, the so-called \textit{experimental design}. In particular, for different trajectories the samples can be taken at different locations, i.e., for $r_1 \neq r_2$ we can have $\ve x^{(r_1, i)} \neq \ve x^{(r_2, i)}$ and in principle even different numbers of samples $N_{r_1} \neq N_{r_2}$. However, here we assume for notational simplicity that $N_r = N$ for all $r$.

\label{sec:stochemu_algorithm}

\begin{figure}[htbp]
	\centering
	\includegraphics[width=.9\textwidth]{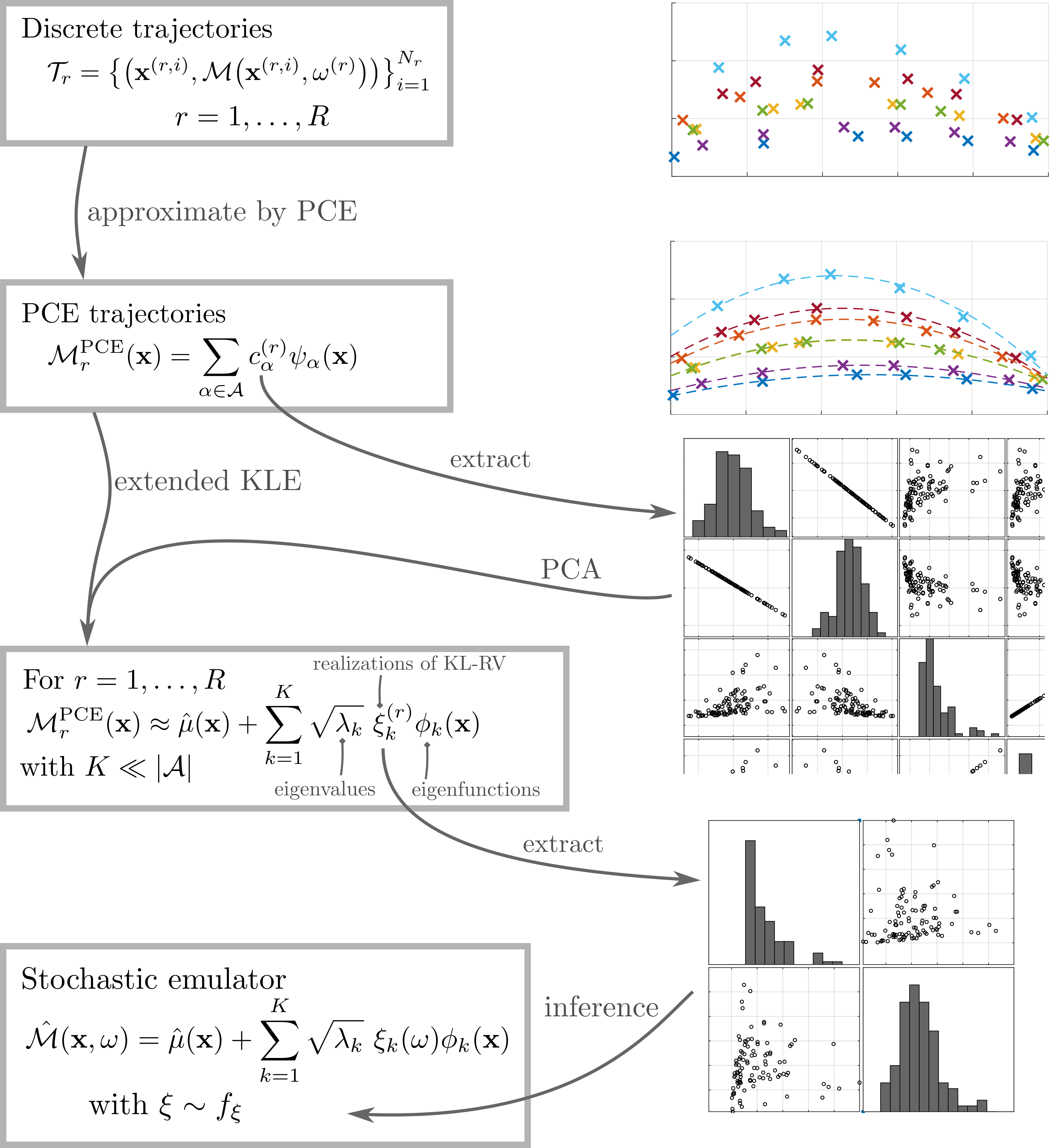}
	\caption{Sketch of our stochastic emulator, starting with stochastic simulator samples (discrete trajectories) at the top and resulting in the stochastic emulator at the bottom, which is a KLE that includes a probabilistic model of the KL-RV. The sketch is purely for illustration and does not display real data. Note that there are two equivalent ways to arrive at the third box: through extended KLE and through PCA on the coefficients.}
	\label{fig:method_sketch}
\end{figure}

Our proposed method consists of the following steps (see also \cref{fig:method_sketch}):
\begin{enumerate}
	
	\item \textbf{Approximate each discrete trajectory} $\ct_r$ by a sparse PCE $\cm^\text{PCE}_r$ in $\ivsp$:
	\begin{equation}
	\cm^\text{PCE}_r(\ve x) = \sum_{\alp \in \ca^{(r)}} \coeff^{(r)}_\alp \psi_\alp(\ve x)
	\end{equation}
	with $\ca^{(r)}$ the set of regressors with nonzero associated coefficient $\coeff^{(r)}_\alp$.
	We use a total-degree basis with degree- and $q$-norm adaptivity to determine the truncation set $\ca^{(r)}$ \citep{BlatmanJCP2011,LuethenIJUQ2022} and apply the least-angle regression solver to compute the coefficients (sparse PCE) \citep{BlatmanJCP2011, LuethenSIAMJUQ2021}.
	
	\item \textbf{Determine a set $\ca$ of regressors} that jointly represents all trajectories well:
	\begin{itemize}
		\item Identify the union $\ca = \bigcup_{r = 1}^{\Nlatent} \ca^{(r)}$ of all chosen regressors.
		\item To keep the size of the basis manageable, discard the regressors with the smallest sum of squares of coefficients over all trajectories ($\sum_{r = 1}^{\Nlatent} \big(\coeff^{(r)}_\alp\big)^2$) until $P = |\ca| \leq \frac{N}{2}$ regressors or less are left in $\ca$.
		\item To avoid discontinuous behavior resulting from sparse selection, recompute the coefficients of every trajectory by ordinary least squares (OLS), using the chosen set of regressors $\ca$.
	\end{itemize}
	This results in $\Nlatent$ PCE trajectories, where each trajectory uses the same set of $P$ PCE basis functions.
	\label{step:PCEtraj}
	
	\item \textbf{Center the PCE trajectories} by subtracting the sample mean
	\begin{equation}
	\hat{\mu}^\text{PCE}(\ve x) = \frac{1}{\Nlatent} \sum_{r = 1}^\Nlatent \cm^\text{PCE}_r(\ve x)
	= \sum_{\alp \in \ca} \left(\frac{1}{\Nlatent} \sum_{r = 1}^\Nlatent \coeff^{(r)}_\alp \right) \psi_\alp(\ve x)
	\label{eq:stochemu_samplemean}
	\end{equation}
	which is itself a PCE. We denote by 
	$\tilde{\cm}^\text{PCE}_r(\ve x) = \cm^\text{PCE}_r(\ve x) - \hat{\mu}^\text{PCE}(\ve x)$ the centered PCE trajectories.
	Extract the coefficients $\tilde{\coeff}^{(r)}_\alp$ of the centered trajectories and store them in a $P\times\Nlatent$ matrix $\tilde{\ve \coeff}$.
	
	\item \textbf{Apply extended KLE} to the set of PCE trajectories. The sample covariance function has the form 
	\begin{equation}
	\hat \covfct(\ve x,\ve x') = \frac{1}{\Nlatent-1} \sum_{r=1}^{\Nlatent} \tilde{\cm}^\text{PCE}_r(\ve x) \tilde{\cm}^\text{PCE}_r(\ve x'). 
	\label{eq:sample_covariance_PCE}
	\end{equation}
	Computing the eigenfunctions $\phi(\ve x)$ of the associated integral eigenvalue problem in \cref{eq:integral_EV_problem} is equivalent to computing a PCA on the PCE coefficients, i.e., equivalent to solving the following $P$-dimensional eigenproblem for $\tilde {\ve a}$: 
	\begin{equation}
	\ve \Sigma \ve \eigv = \lambda \ve \eigv,
	\end{equation}
	where $\ve\Sigma = \frac{1}{\Nlatent-1} \tilde {\ve \coeff} \tilde{\ve \coeff}^T$
	(see \cref{app:derivations_KLE_PCE} for the derivation of this equivalence).
	The eigenvectors $\ve \eigv$ contain the coefficients of the eigenfunctions represented in the PCE basis: $\phi(\ve x) = \sum_{\alp \in \ca} \eigv_\alp \psi_\alp(\ve x)$.

	\item \textbf{Identify the truncation order} $K \ll P$ for the KLE based on a given threshold for the explained variance. We use a threshold of $99.9\%$ (see \cref{eq:KLEthreshold}).
	\label{step:truncation}
	
	\item \textbf{Compute the realizations} of the KL-RV $\xi_i$ from the sample trajectories by projecting onto the eigenfunctions. Due to the orthonormality of the PCE basis, this can be done analytically (see \cref{app:derivations_KL-RV}). Denote the realizations by $\ve\xi^{(r)} \in \Rr^K$.
	\label{step:KLRVrealizations}
	
	\item \textbf{Infer the joint distribution} $f_{\ve \xi}$ of random KL coefficients from the data set 
	$\{\ve\xi^{(r)}\}_{r = 1 \enum \Nlatent}$. 
	We will test four methods consisting of the techniques described in \cref{sec:inference_distributions}:
	\label{step:KLRVinference}
	\begin{enumerate}
		\item Option 1: assume standard Gaussian marginals, which implies independence;
		\item Option 2: parametric inference of the marginals (with moment constraints) and of the copula;
		\label{step:KLinf_paramEmu}
		\item Option 3: 1D kernel density estimation of each marginal, assuming independence;
		\item Option 4: 1D kernel density estimation of each marginal and parametric inference of the copula.
	\end{enumerate}

\end{enumerate}

The resulting stochastic model for the random field $\cm$ is 
\begin{equation}
\hat\cm(\ve x, \cdot) = \hat{\mu}(\ve x) + \sum_{k=1}^K \sqrt{\lambda_k} \ Z_k(\cdot) 
\underbrace{\left( \sum_{\alp \in \ca} \eigv^{(k)}_\alp \psi_\alp(\ve x) \right)}_{=\phi_k(\ve x)}
\label{eq:stochastic_emulator}
\end{equation}
where $\ve Z = (Z_1, Z_2 \enum Z_K)$ is a random vector distributed according to the inferred joint distribution $f_{\ve \xi}$.

The full procedure is visualized in \cref{fig:method_sketch}.

Using the stochastic emulator constructed in \cref{eq:stochastic_emulator}, we can easily compute the following quantities.
\begin{itemize}
	\item The mean function $\hat \mu$ is given by the sample mean of the approximated trajectories (PCE trajectories), see \cref{eq:stochemu_samplemean}.
	
	\item The covariance function $\hat \covfct(\cdot, \cdot)$ can be computed from the KLE eigenfunctions using the truncated version of \cref{eq:covfct_series}.
	Note that this relation does not involve the KL-RV.

	\item New trajectories (i.e., realizations of the random field) can be generated by drawing new samples of the KL-RV $\xi_k$, and evaluating \cref{eq:KLE}. 
	
	\item A histogram of the marginal pdf $f_{\cm_{\ve x'}}$ of the random field at any input space location $\ve x'$ can be created by generating many new trajectories and evaluating them at $\ve x'$.
	
\end{itemize}

\begin{remark}[Another stochastic emulator] 
	\label{remark:PCE-KDE}
	A simple stochastic emulator able to model marginal distributions $f_{\cm_{\ve x'}}$ can be constructed by 
	evaluating all PCE trajectories from Step~\ref{step:PCEtraj} above at the new location $\ve x'$ and computing a kernel density estimate on the resulting set of predictions. This method will be used as a comparison method for marginal estimation in \cref{sec:numerical_experiments}. 
	However, unlike our stochastic emulator in \cref{eq:stochastic_emulator}, this simple emulator is not able to resample trajectories. 
\end{remark}

\begin{remark}[Alternatives to PCE] 
	We choose PCE to approximate the sampled trajectories because it is a powerful method for deterministic surrogate modeling. However, the choice of PCE in the above method is not crucial: without any changes to the methodology, PCE could be replaced by any other spectral expansion onto an orthonormal basis of $\ivsp$, e.g., a Poincar\'e basis \citep{LuethenPoincare2022} or a spline basis \citep{Rahman2020}.
	From the orthonormality of the basis it follows that functional PCA in $\ivsp$ becomes traditional (unweighted) PCA in the coefficient space (see \cref{app:derivations_KLE_PCE}), which avoids the expensive numerical solution of the integral eigenvalue problem in $d$ dimensions, and instead solves an inexpensive  discrete eigenvalue problem.
\end{remark}

\FloatBarrier

\section{Numerical experiments}
\label{sec:numerical_experiments}
\label{sec:error_measures}

To analyse the performance of our stochastic emulator, we apply it to three models of increasing complexity: the three-dimensional Ishigami function with two random parameters (\cref{sec:ishigami}), the borehole model with five hidden (latent) variables (\cref{sec:borehole}), and finally the Heston stochastic volatility model, a system of two stochastic ODEs with six inputs that has already been used by \citet{ZhuRESS2021} as a stochastic emulator benchmark model (\cref{sec:heston}).

We first investigate the pointwise approximation capabilities of our emulator by plotting the stochastic simulator and emulator responses at selected points throughout the input domain. Then, we investigate the convergence behavior of our stochastic emulator using the following global error measures:
\begin{itemize}
	
	\item The global convergence of the marginal distributions is assessed using the \textit{averaged normalized Wasserstein distance}.
	The \textit{Wasserstein distance of order two} between two random variables $Y_1$, $Y_2$ with quantile functions (inverse CDF) $Q_1$, $Q_2$ is defined by \citep{Villani2009} 
	\begin{equation}
	d_\text{WS}(Y_1, Y_2) = \norme{Q_1 - Q_2}{2} = \sqrt{\int_0^1 (Q_1(u) - Q_2(u))^2 \ \di{u}}.
	\end{equation}
	To measure the global quality of marginal approximation, we consider the quantity
	\begin{equation}
	\epsilon_{\text{marg}} = \Espe{\ve X}{\frac{d_\text{WS} \big(\cm(\ve X, \cdot), \hat \cm(\ve X, \cdot) \big)}{\sigma(\cm(\ve X, \cdot))}},
	\label{eq:wasserstein_avno}
	\end{equation}
	computed by Monte-Carlo integration on a validation set with $\Nparams_\text{val} = 1,000$ points and $\Nlatent_\text{val} = 10,000$ replications \citep{ZhuSIAM2021}.
	
	\item The global error between the true covariance function $\covfct$ and the emulated one $\hat{\covfct}$ is computed by
	\begin{equation}
	\epsilon_\text{cov} = \norme{\covfct - \hat \covfct}{\ivsp\times \ivsp} 
	\approx 
	\frac{1}{\Nparams_\text{val}}\norme{C - \hat C}{F}
	\label{eq:error_covariance}
	\end{equation}
	where $C$ and $\hat C$ denote the true and emulated covariance matrices for a validation sample $\{\ve x^{(i)}: i = 1 \enum \Nparams_\text{val}\}$, and $\norme{\cdot}{F}$ is the Frobenius norm.
	
\end{itemize}

\subsection{Stochastic Ishigami function}
\label{sec:ishigami}

\subsubsection{Problem statement}

\begin{figure}[htb]
	\subfloat[Input space with validation points used for visualization]
	{
		\includegraphics[width=.49\textwidth, trim=.8cm 0cm 1.2cm 1cm, clip]
		{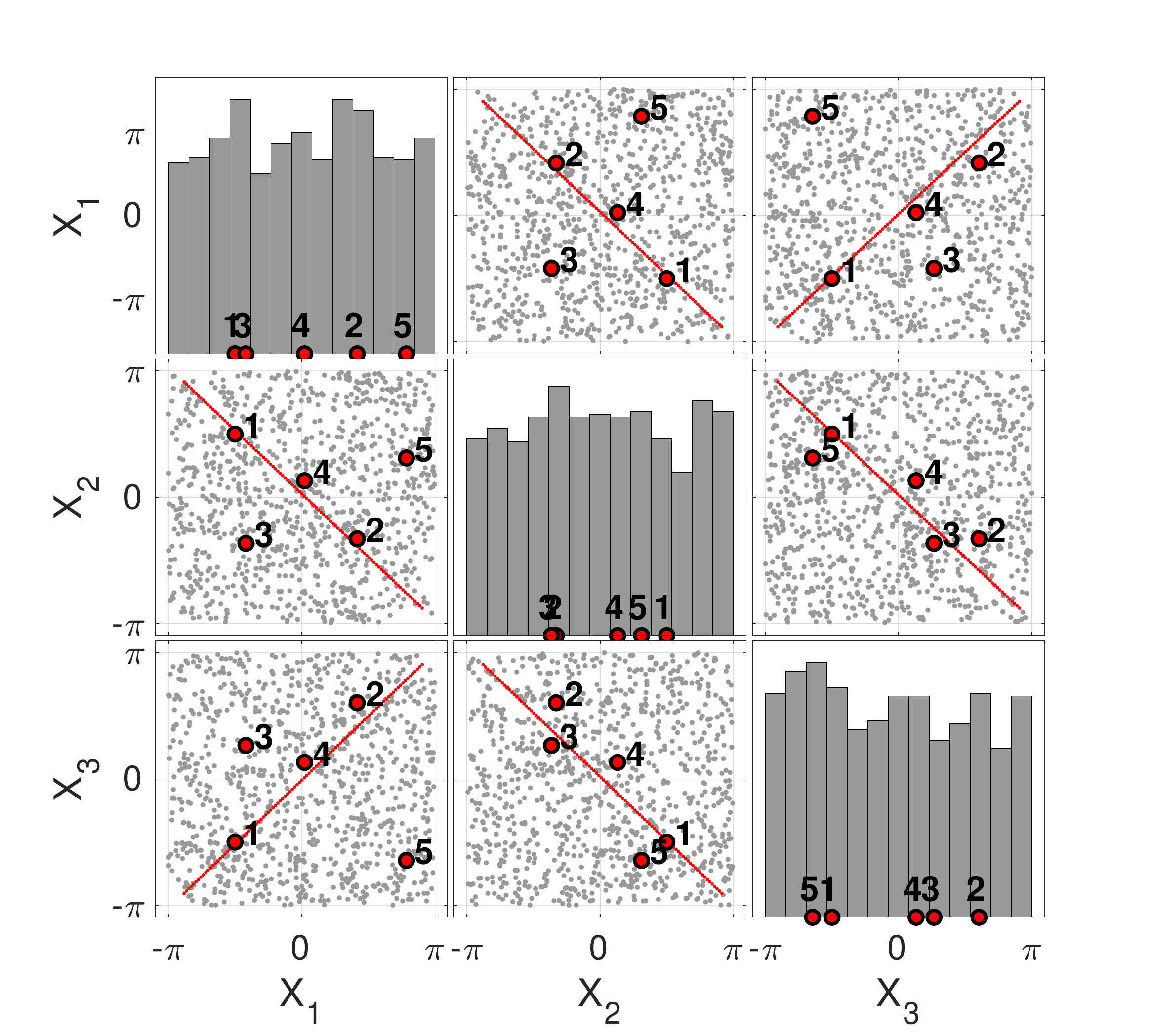}
		\label{fig:ishigami_input}
	}%
	\subfloat[Latent space]
	{
		\includegraphics[width=.49\textwidth, trim=.8cm 0cm 1.2cm 1cm, clip]{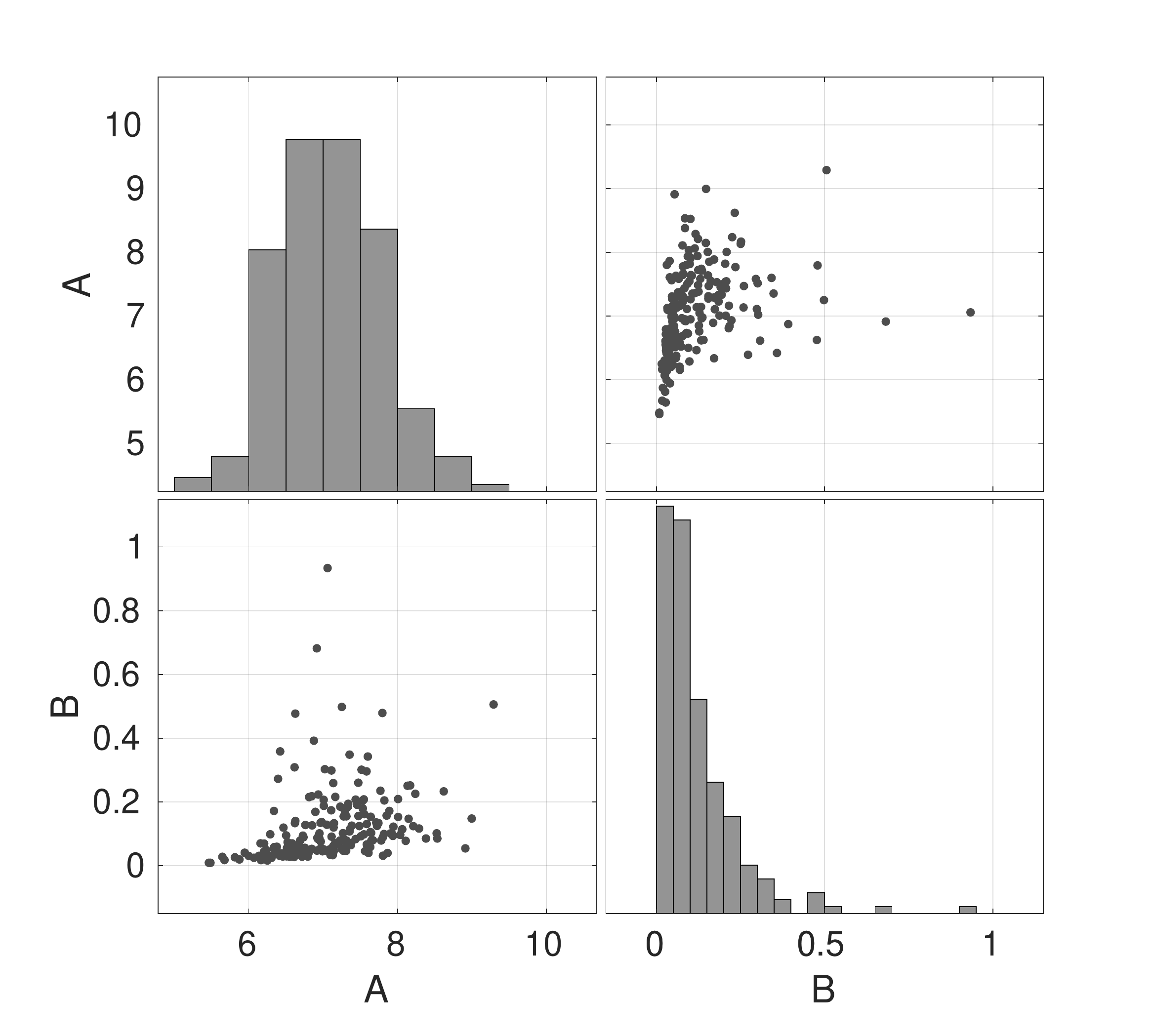}
		\label{fig:ishigami_latentspace}
	}
	\caption{
		Samples of the input space (left) and the latent space (right). The red line in \cref{fig:ishigami_input} is the trajectory along which the simulator/emulator response is plotted in \cref{fig:ishigami_1Dtrajectory}. The red dots annotated by small numbers denote the five points that are used for visualization in the following. \cref{fig:ishigami_latentspace} shows a sample of the latent space ($A, B$ in \cref{eq:ishigamimodel}).}
	\label{fig:ishigami_input_latent}
\end{figure}

The Ishigami function is a a well-known benchmark function for deterministic surrogate models. It is highly non-linear and has significant interaction terms. 
It becomes a stochastic simulator by treating its parameters $a$ and $b$, which are usually fixed at $a = 0.7$ and $b = 0.1$, as additional random variables:
\begin{equation}
f(\ve X; A, B) = \sin(X_1) + A \sin(X_2)^2 + B X_3^4 \sin(X_1).
\label{eq:ishigamimodel}
\end{equation}
$A$ and $B$ have here the role of so-called \textit{hidden} or \textit{latent} random variables. In other words, we assume that they cannot be observed, and that therefore their values cannot be utilized in the surrogate modeling process. 
They introduce stochasticity into the otherwise deterministic Ishigami model. 
Here we model $A$ and $B$ as lognormal random variables with mean 7 and standard deviation 0.7, and mean 0.1 and standard deviation 0.1, respectively. We assume that both variables are coupled with a Clayton pair copula with parameter 1.5.
The non-hidden (explicit) input variables are as usual $\ve X = (X_1, X_2, X_3)$, which are independent and uniformly distributed in $[-\pi, \pi]$.
A Sobol' analysis of $f(\ve X; A, B)$ in \cref{eq:ishigamimodel} reveals that the main effect of the group of explicit input parameters ($\cm_1$ in \cref{eq:stochsim_decomp}) is approx.\ $75\%$, while the interaction effect between the explicit and the latent group is approx.\ $25\%$, and the main effect of the group of latent variables is negligible. 
Samples of the input space and the latent space are displayed in \cref{fig:ishigami_input_latent}.

We use different experimental design sizes $\Nparam \in \{50, 100, 150\}$ and a maximum degree of $p = 14$ for the PCE trajectories (with degree-adaptivity \citep{BlatmanJCP2011}). This results in a relative mean-squared error in the order of $10^{-3}/ 10^{-5}/ 10^{-10}$, respectively.
We also test different numbers of trajectories $\Nlatent \in \{10, 30, 100, 300\}$,
and use a different experimental design for each trajectory.
For each combination of experimental design size and number of trajectories, we conduct 50 independent repetitions. All resulting stochastic emulators are evaluated on the same validation set, consisting of $\Nlatent_\text{val} = 10,000$ trajectories of the true stochastic simulator, each evaluated on a set of $\Nparams_\text{val} = 1,000$ points in the input space.

\subsubsection{Analysis of the KL-RV samples}
To illustrate the type of result obtained with our proposed stochastic emulator  described in \cref{sec:our_approach}, we now present scatterplots showing realizations of the following random quantities: 1) the KL-RV (compressed representation of PCE coefficients) resulting from step \ref{step:KLRVinference}; 2) the PCE coefficients resulting from transforming the KL-RV samples to the PCE coefficient space.
Detailed results for the prediction $\cm(\ve x', \cdot)$ at a new location $\ve x'$ for a number of new trajectories are presented in \cref{sec:ishigami_marginals} below.
The results shown here are based on parametric inference of marginals and copula (Option (b) of Step~\ref{step:KLRVinference}).

\begin{figure}[htb]
	\centering
	\includegraphics[width=.6\textwidth, trim=1cm .2cm .5cm 1cm, clip]
	{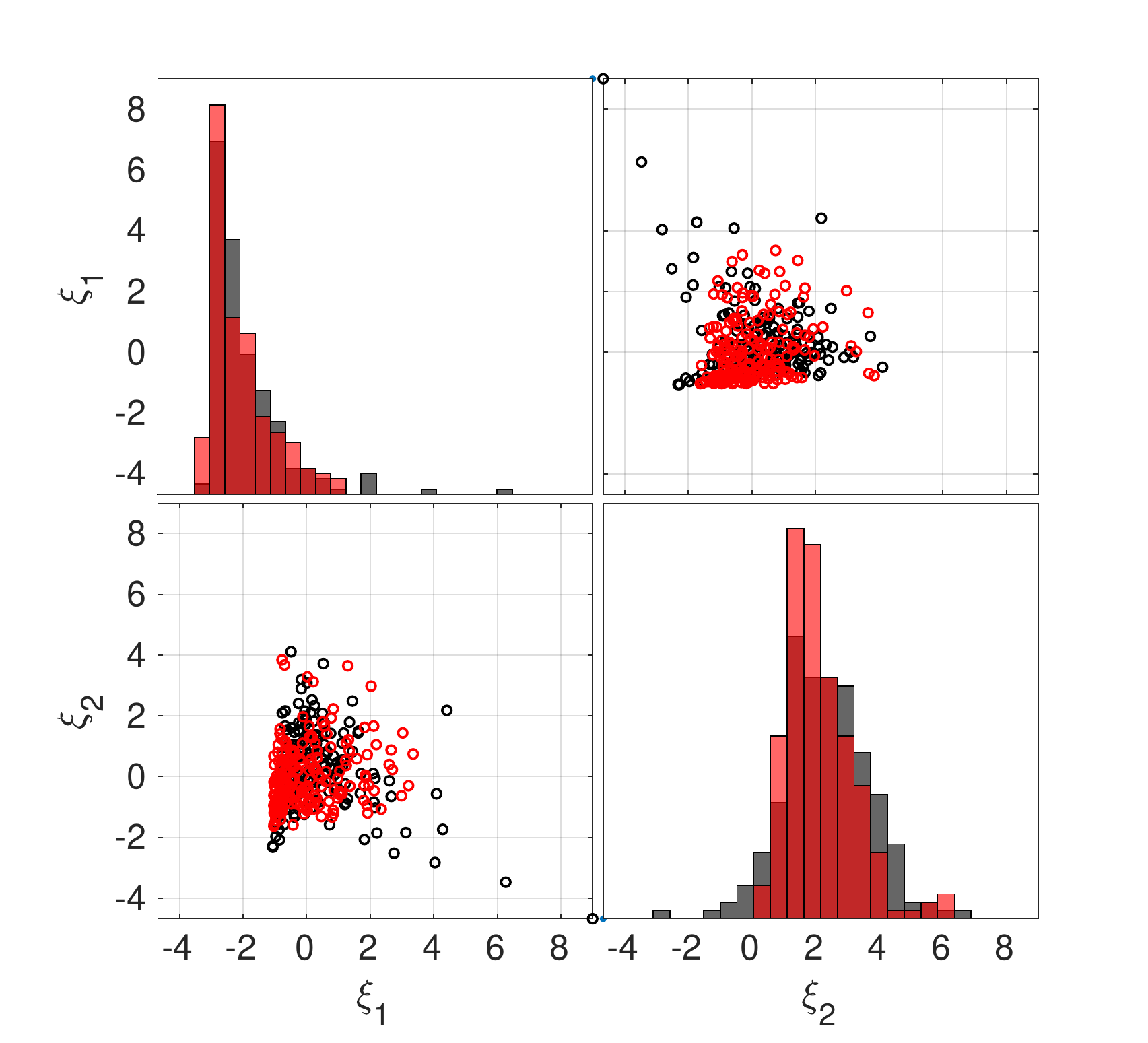}
	\caption{KL-RV coefficient samples computed from validation trajectories (black) and new samples from the stochastic emulator (red). This is data from one experiment with $\Nparams = 150$ and $\Nlatent = 100$, max degree $p = 14$. Number of validation trajectories and resampled PCE coefficients: $200$ each. Number of KL modes: $\Nkl = 2$. Inferred distribution of KL-RV: Beta and Gumbel, with Clayton copula (parameter $0.32$). The corresponding eigenvalues are $\lambda_1 = 4.46$ and $\lambda_2 = 0.10$.}
	\label{fig:ishigami_KLRVsample}	
\end{figure}

The truncation of the KLE (Step~\ref{step:truncation} of our algorithm) typically results in two modes with eigenvalues $\lambda_1 \in [3,5]$ and $\lambda_2 \approx 0.1$ depending on the size and realization of the experimental design.
We display a specific example in \cref{fig:ishigami_KLRVsample}, which is computed from 100 trajectories with 150 samples each, and has eigenvalues $\lambda_1 = 4.46$ and $\lambda_2 = 0.10$. 
The figure shows resampled values for the KL-RV: in black, samples computed from validation trajectories by projecting first onto the truncated PCE space and then onto the eigenfunctions; in red, new samples drawn from the input object inferred in Step~\ref{step:KLRVinference} of our algorithm (\cref{sec:stochemu_algorithm}). We see that their inferred joint distribution (Beta and Gumbel marginals, with a Clayton copula) visually matches the validation data well.

\begin{figure}[htb]
	\centering
	\includegraphics[width=.8\textwidth, trim=1cm 1cm 2cm 1.8cm, clip]
	{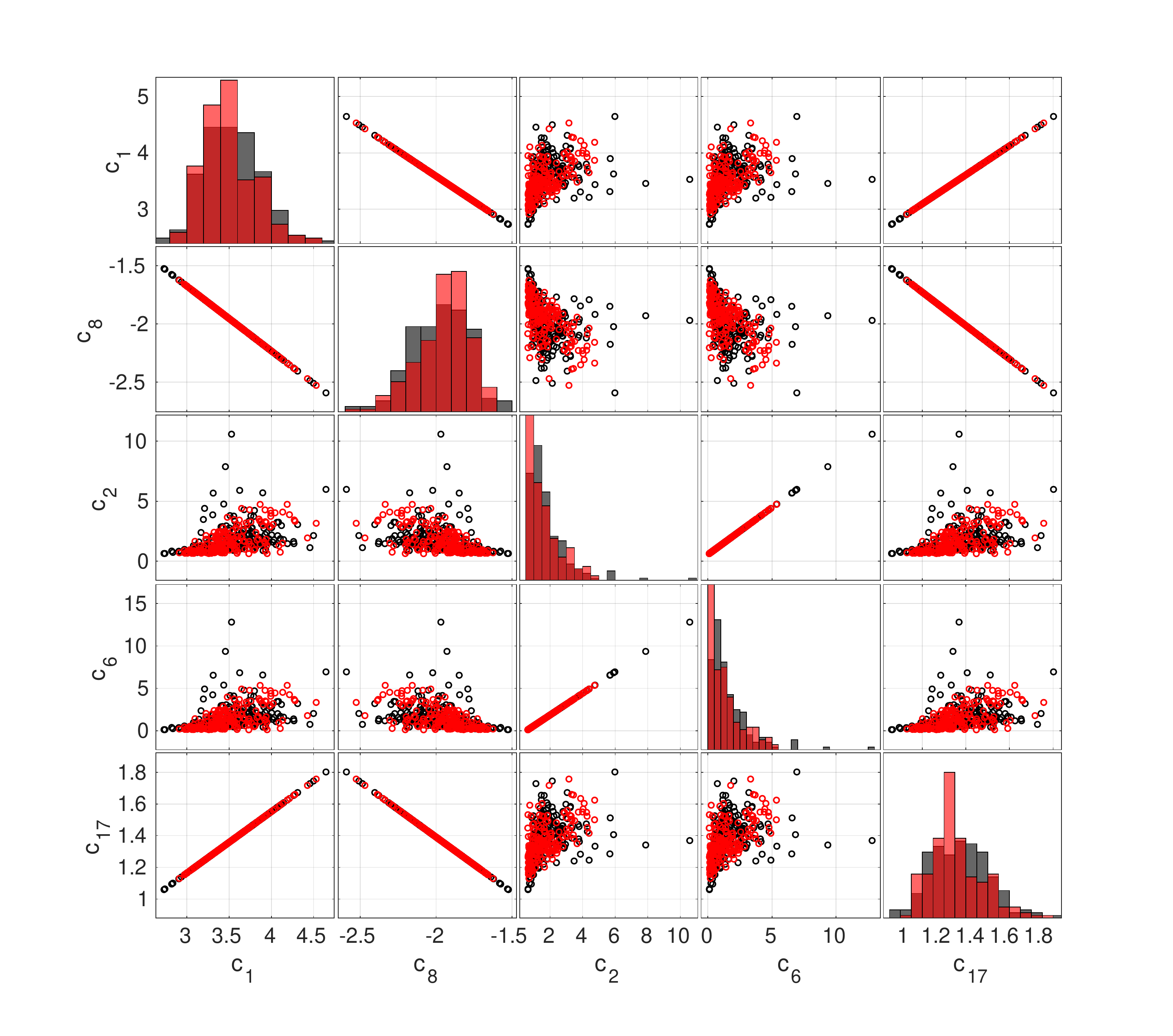}
	\caption{PCE coefficient samples computed from validation trajectories (black) and new samples from the stochastic emulator (red). This is data from one experiment with $\Nparams = 120$ and $\Nlatent = 100$, max degree $p = 14$. Number of validation trajectories and resampled PCE coefficients: 
		$\Nlatent_\text{val} = 200$ each. 
		The PCE coefficients are sorted by mean magnitude, and we only display the largest 5 out of {total 75 nonzero} coefficients.}
	\label{fig:ishigami_PCEsample}	
\end{figure}

The KL-RV are the compressed representation of the random PCE coefficients (which in turn encode trajectories).
Mapping the realizations of the KL-RV back to the PCE coefficient space, we obtain the samples 
displayed in \cref{fig:ishigami_PCEsample} for $\Nparams = 150$ and $\Nlatent = 100$.
Validation samples from the original stochastic simulator (generated by regressing them onto the truncated PCE basis) are displayed in black, while 200 resampled PCE coefficient vectors generated from the stochastic emulator are shown in red. We only show the 5 coefficients with maximal mean absolute value. 
We see that the validation samples have a slightly larger spread than the emulator samples, but that overall the behavior is matched well. 
Some parameters have linear functional dependence, e.g., $\coeff_1, \coeff_8$ and $\coeff_{17}$, which is perfectly reproduced by the emulator. These parameters correspond to the basis functions $\alp_1 = (0,0,0)$ (constant term), $\alp_8 = (0,4,0)$ and $\alp_{17} = (0,6,0)$ and are needed to emulate the second term of the stochastic Ishigami model in \cref{eq:ishigamimodel}. There are no interactions with the other terms, therefore a different value of $A$ just proportionally changes the relative weighting of these terms. 
A similar explanation holds for $\coeff_2$ and $\coeff_6$ with $\alp_2 = (1,0,0)$ and $\alp_6 = (1,0,2)$, which are involved in emulating the first and the third term of \cref{eq:ishigamimodel} and change proportionally with $B$.

\subsubsection{Marginal performance on selected validation points}
\label{sec:ishigami_marginals}

We now investigate the performance of the stochastic emulator with parametric inference of KL-RV marginals and copulas (choice \ref{step:KLinf_paramEmu}) on a selection of out-of-sample validation points, i.e., points that were not used for training.

\begin{figure}[htb]
	\centering
	\includegraphics[width=.9\textwidth, trim=1cm 1.2cm 2cm .5cm, clip]
	{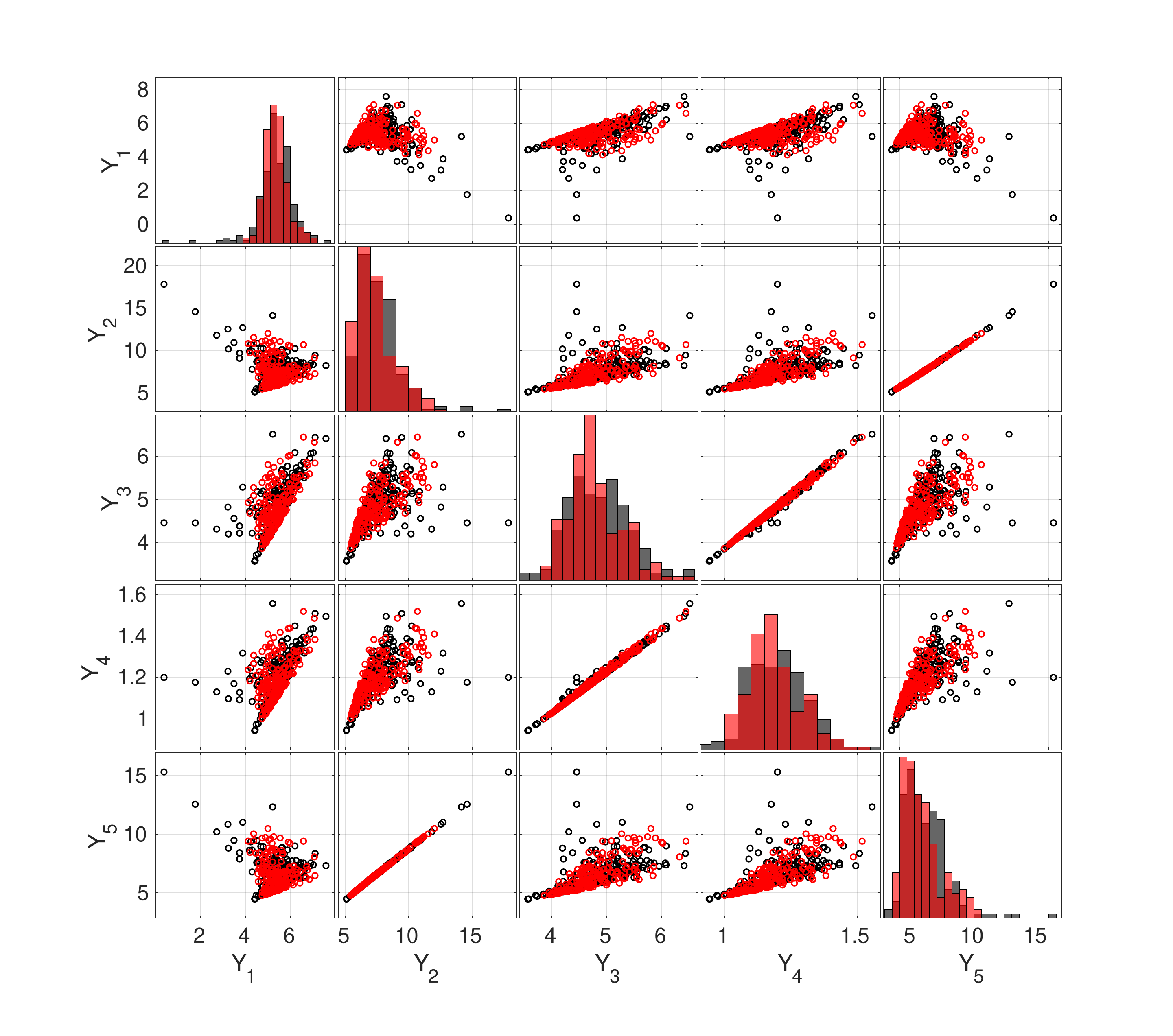}
	\caption{Scatterplot of output $Y_i = \cm(\ve x^{(i)}, \cdot)$ of the stochastic simulator (black) and of output $\hat Y_{i} = \hat\cm(\ve x^{(i)}, \cdot)$ of the parametric stochastic emulator (red, created from training set with $\Nparams=150$, $\Nlatent = 100$) for five validation points sampled from the input space. The location of these five points is illustrated in \cref{fig:ishigami_input}. }
	\label{fig:ishigami_Ypred}
\end{figure}

\cref{fig:ishigami_Ypred} shows the histograms and pairwise scatterplots of samples from the output random variables $Y_{i} = \cm(\ve x^{(i)}, \cdot)$ and $\hat Y_{i} = \hat\cm(\ve x^{(i)}, \cdot)$ of the stochastic simulator and emulator, respectively. The five selected validation locations $\{\ve x^{(i)}\}_{i=1}^5$ in the input space are visualized in \cref{fig:ishigami_input} by red dots. Each black (resp.~red) point in \cref{fig:ishigami_Ypred} is a new trajectory of the stochastic simulator (resp.~emulator) evaluated at the five given points. Both samples have the same size (200 new trajectories). Overall, the model behavior is captured well, but the stochastic simulator has a slightly larger spread (see e.g.\ $Y_2$ vs.\ $Y_3$).

\begin{figure}[htb]
	\centering
	{\includegraphics[width=.9\textwidth]{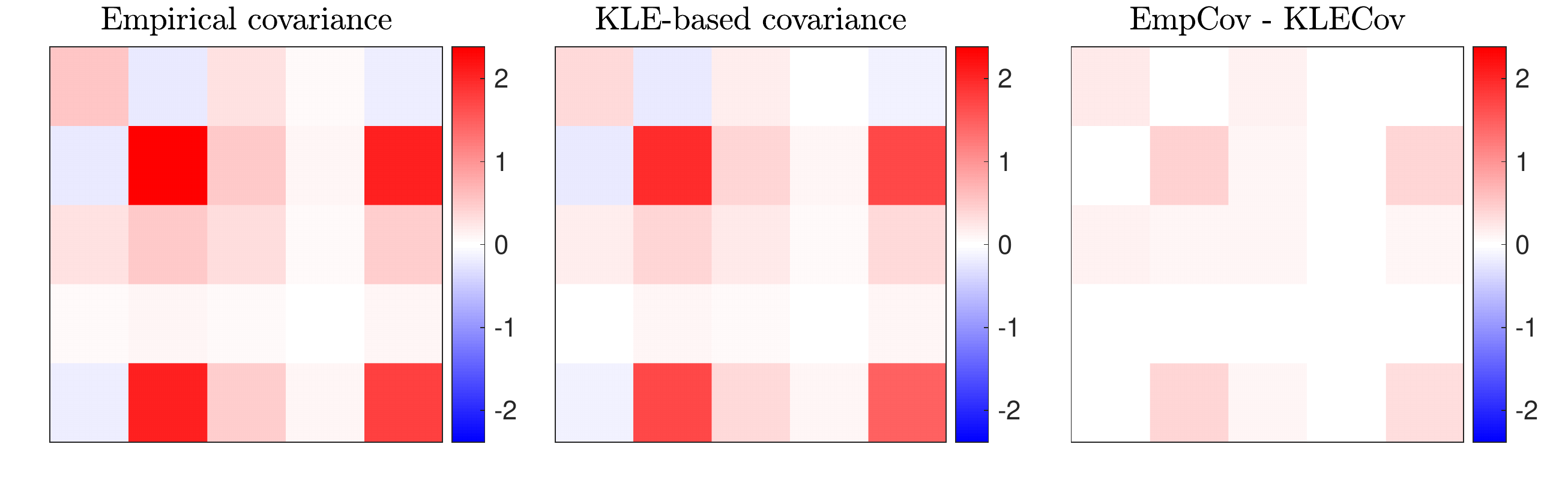}}
	\caption{
		Covariance approximation for the Ishigami model for the $N_\text{val}=5$ validation locations in the input space illustrated in \cref{fig:ishigami_input}. Computation based on $\Nparams = 150, \Nlatent = 100$, max degree $p = 14$. 
		KLE-based covariance: computed from eigenfunctions as in \cref{eq:covfct_series}.
		Empirical covariance: based on the validation set comprising $10,000$ trajectories of the stochastic simulator.}
	\label{fig:ishigami_vis_covariance}	 
\end{figure}

From the data in the off-diagonal scatter plots in \cref{fig:ishigami_Ypred}, we can compute the sample covariance matrix. However, we can also compute the covariance analytically from the KLE eigenfunctions, using \cref{eq:covfct_series}. 
In \cref{fig:ishigami_vis_covariance}, we use the five illustrative points shown in \cref{fig:ishigami_input} to compare this covariance estimate to a validation covariance matrix computed empirically from $10,000$ trajectories of the stochastic simulator.
Qualitatively, the covariance is reproduced well, although the KLE-based covariance is slightly smaller in magnitude than the empirical covariance. 

\begin{figure}[htbp]
	\centering
	\subfloat[][$\Nlatent = 10$]{\includegraphics[width=.245\textwidth, trim=.3cm .1cm .6cm .4cm, clip]{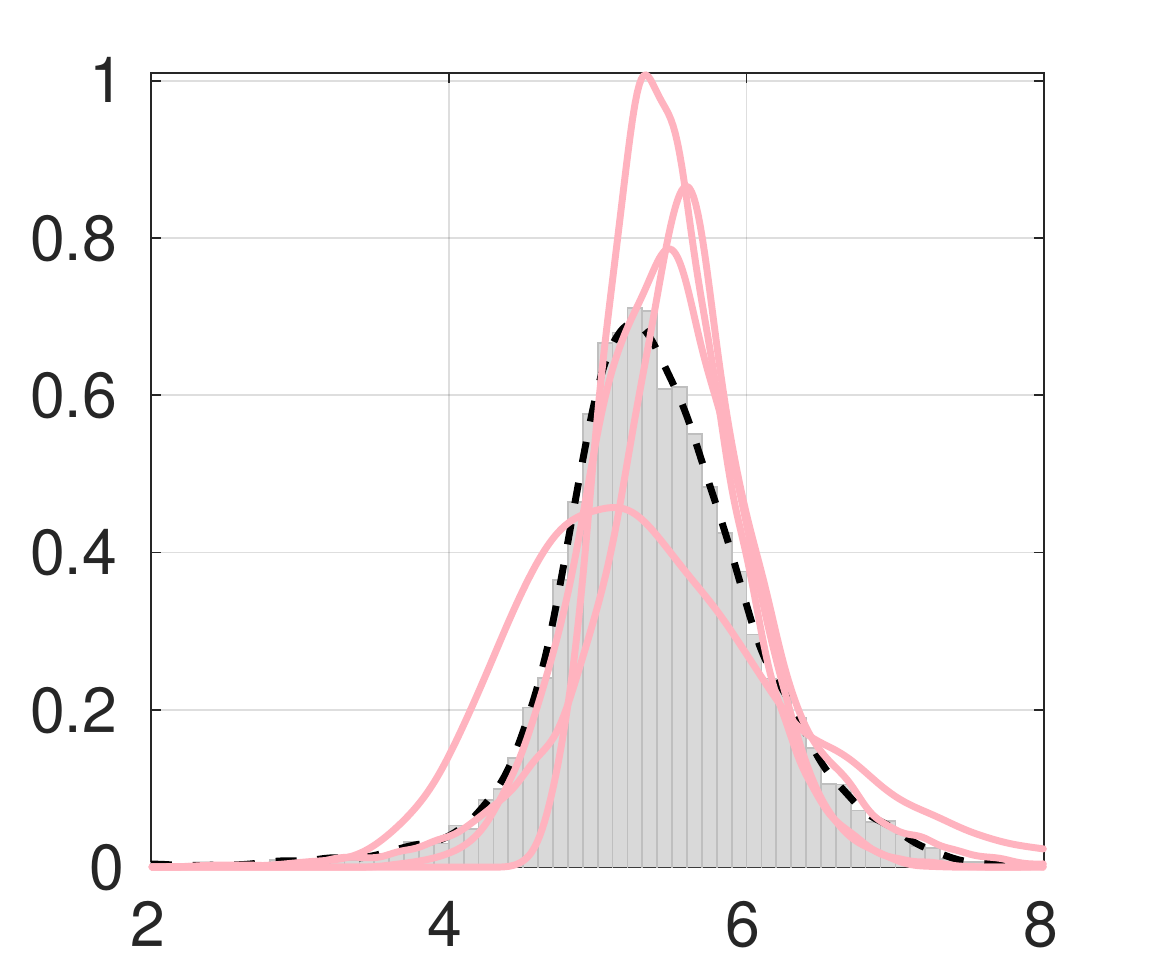}}
	\hfill
	\subfloat[][$\Nlatent = 30$]{\includegraphics[width=.245\textwidth, trim=.3cm .1cm .6cm .4cm, clip]{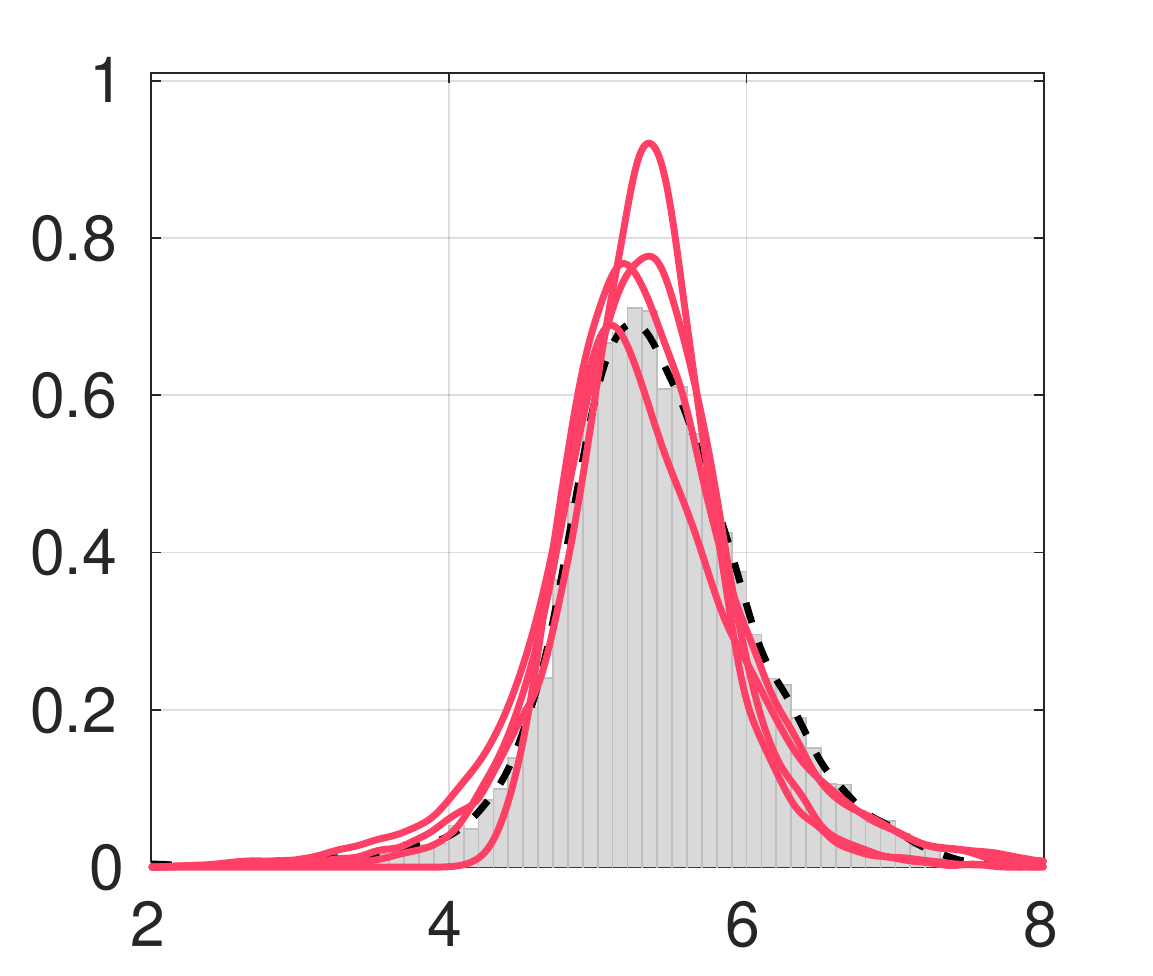}}
	\hfill
	\subfloat[][$\Nlatent = 100$]{\includegraphics[width=.245\textwidth, trim=.3cm .1cm .6cm .4cm, clip]{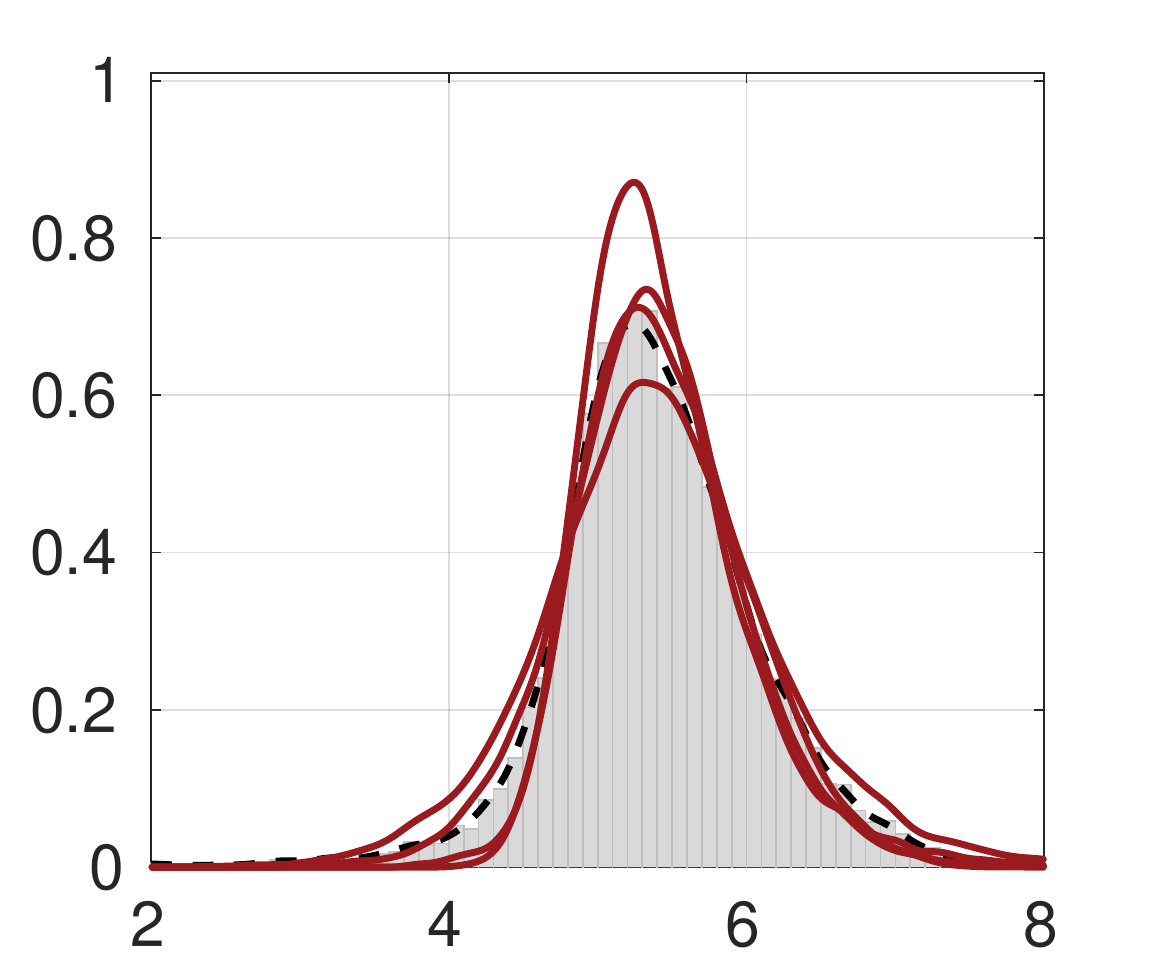}}
	\hfill
	\subfloat[][$\Nlatent = 300$]{\includegraphics[width=.245\textwidth, trim=.3cm .1cm .6cm .4cm, clip]{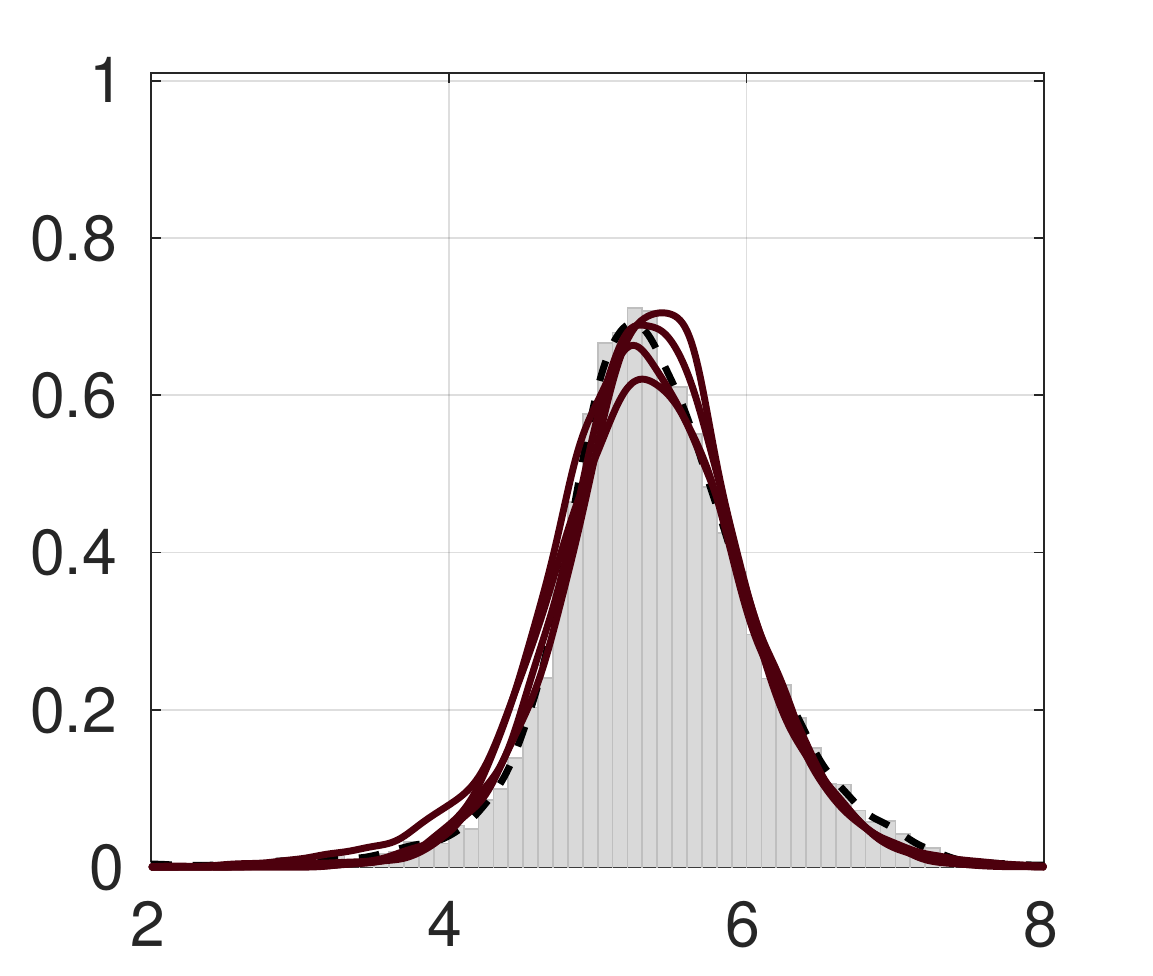}}
	\caption{Prediction of $Y$-marginal at validation point $\ve x_\text{val}^{(1)} = (-\frac{\pi}{2}, \frac{\pi}{2}, -\frac{\pi}{2})$ for the Ishigami model and $\Nlatent = \{10, 30, 100, 300\}$ trajectories with the parametric stochastic emulator for 4 independent repetitions. Visualization of predicted marginals by KDE using $10,000$ samples. Number of experimental design points $\Nparams = 150$, max degree $p = 14$. {The approximation error is in the order $\co(10^{-10})$.} 
	}
	\label{fig:ishigami_vis_marginals}	
\end{figure}

In \cref{fig:ishigami_vis_marginals}, we visualize the marginal distribution $f_{Y_1}$ of $Y_{1} = \cm(\ve x^{(1)}, \cdot)$ at one validation point (the point marked with ``1'' in \cref{fig:ishigami_input}) for an increasing number of trajectories in the training set, and 4 independent repetitions of each experiment. 
The estimates for the marginal distribution $f_{Y_1}$ are computed by KDE from $10,000$ samples from the constructed stochastic emulator, while the histogram and the dashed curve represents a validation set of $10,000$ samples of the original stochastic simulator.
As expected, we observe that with an increasing number $\Nlatent$ of trajectories, the shape of the predicted marginal becomes closer to the kernel density estimate of the validation set and shows less variation.

\begin{figure}[htb]
	\centering
	\subfloat[Stochastic simulator]{\includegraphics[height=4.5cm, keepaspectratio, trim=.5cm 1.2cm 30.4cm .7cm, clip]
		{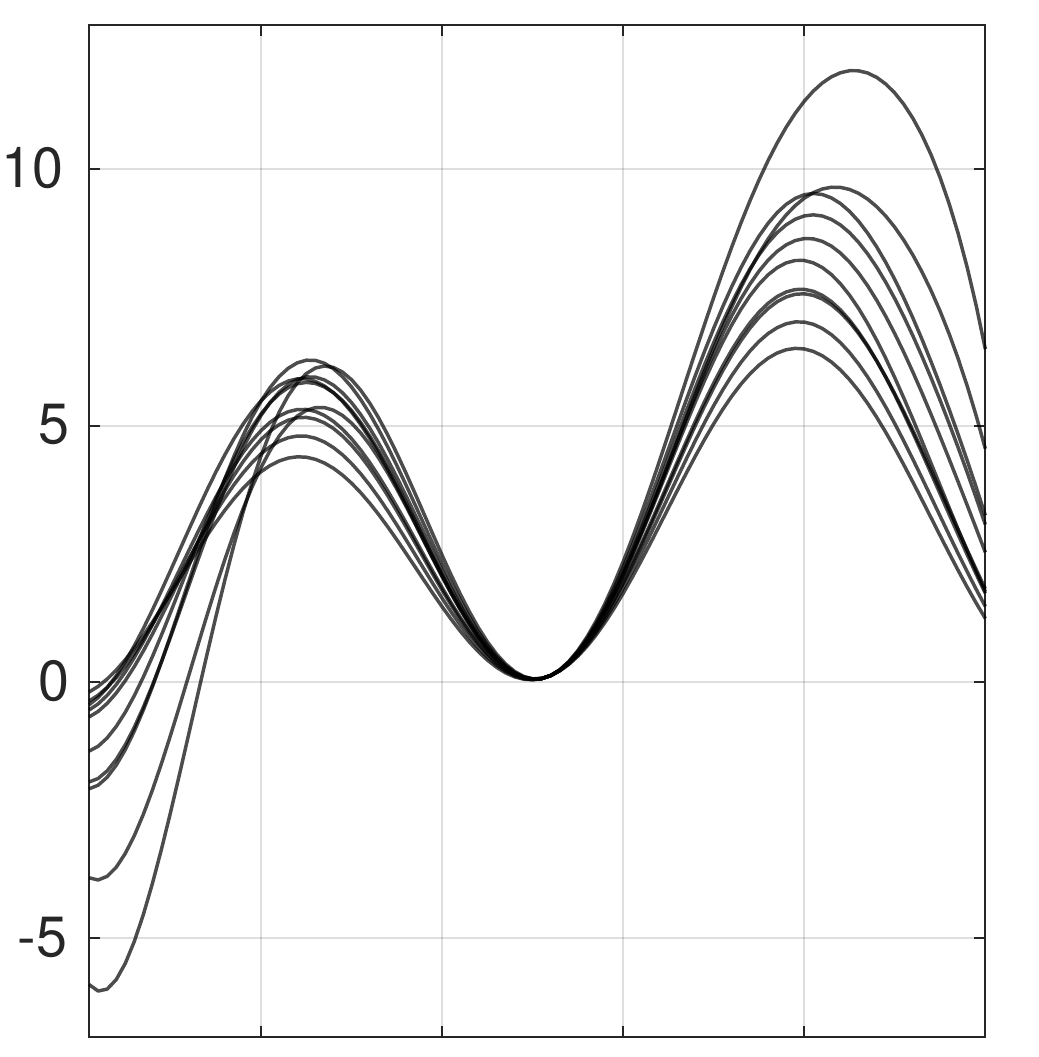}}%
	\subfloat[Stochastic emulator]{\includegraphics[height=4.5cm, keepaspectratio, trim=11cm 1.2cm 20cm .7cm, clip]
		{Figures/ishigami/trajectory1D_N150_R100.pdf}}%
	\subfloat[Comparison of quantiles for both models]{\includegraphics[height=4.5cm, keepaspectratio, trim=22cm 1.2cm 2cm .7cm, clip]
		{Figures/ishigami/trajectory1D_N150_R100.pdf}\label{fig:ishigami_1Dtrajectory_quantiles}}
	\caption{Visualization of stochastic simulator/emulator response $Y$ when following a 1D slice through the input space, which is illustrated with a red line in \cref{fig:ishigami_input}. $\Nparams = 150, \Nlatent = 100$. The left and middle plots show 10 trajectories each. The right plot aggregates the values from $10,000$ trajectories to show quantile information.}
	\label{fig:ishigami_1Dtrajectory}
\end{figure}

Finally, to assess visually how well the resampled trajectories match the behavior of the original stochastic model, we plot in \cref{fig:ishigami_1Dtrajectory} a 1D slice of 10 new trajectories generated by the stochastic simulator (left) and the stochastic emulator (middle). 
The slice through the input space is shown in \cref{fig:ishigami_input} by a red line. On the right, data for the same slice is shown, but this time we show quantiles aggregated over $10,000$ new trajectories each. The trajectory slices look qualitatively similar, although there is a lot of variability between individual realizations. 
From the aggregated data on the left, we see 
that the bulk of the distribution (10\%-90\% quantile) is predicted quite accurately.
Interestingly, in \cref{fig:ishigami_1Dtrajectory_quantiles} it seems that the trajectories of the stochastic emulator (red) have a larger spread than the ones of the simulator (black), contrary to the results earlier in this section, which always showed the simulator having a larger spread than the emulator. This illustrates the difficulty of inferring global behavior from local observations. Theoretically, the emulator should have a smaller variance than the simulator, because terms are missing from \cref{eq:covfct_series} due to truncation.

\subsubsection{Convergence with the number of trajectories}
\label{sec:ishigami_convergence}

To assess the global performance of our proposed method, we now construct stochastic emulators for all combinations of input space experimental design sizes $\Nparam \in \{50, 100, 150\}$ and numbers of trajectories in the range $\Nlatent \in \{10, 30, 100, 300\}$. 
We then evaluate each of the resulting stochastic emulators $\Nlatent_\text{val} = 10,000$ times at $\Nparams_\text{val} = 1,000$ validation points in the input space (out-of-sample, i.e. not used for training)
and compute the errors as described in \cref{sec:numerical_experiments}. 
Each combination is independently repeated 50 times to account for the statistical uncertainty of the sampling of both experimental design and trajectories, which allows us to display results in the form of Tukey boxplots.

\begin{figure}[htb]
	\centering
	\subfloat[Convergence of $\epsilon_\text{marg}$]
	{\includegraphics[width=.7\textwidth, trim=.2cm 0 1.4cm .1cm, clip, height=.19\textheight, keepaspectratio]
		{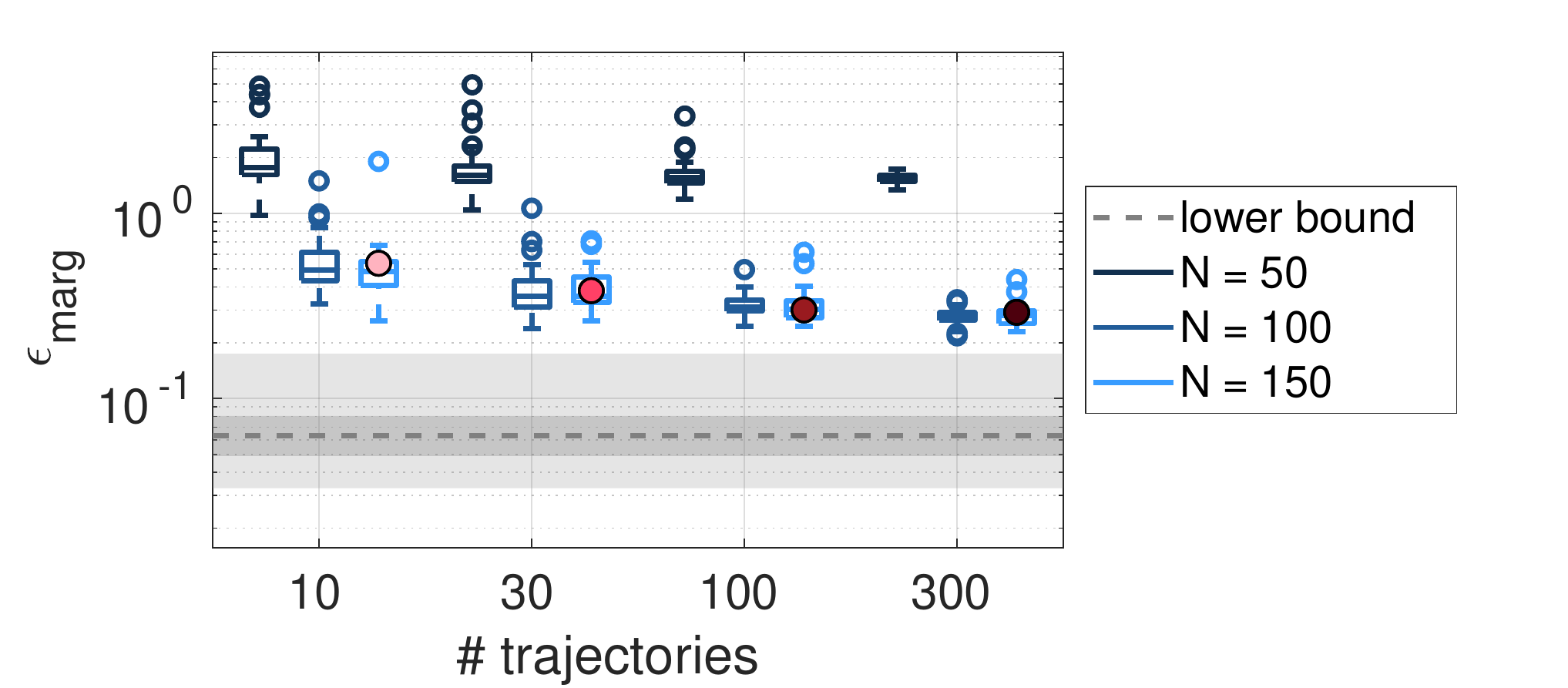}
		\label{fig:ishigami_convergence_marginals}}
	\hfill	
	\subfloat[Marginal at $\ve x_\text{val}^{(1)} = (-\frac{\pi}{2}, \frac{\pi}{2}, -\frac{\pi}{2})$]
	{\includegraphics[width=.35\textwidth]
		{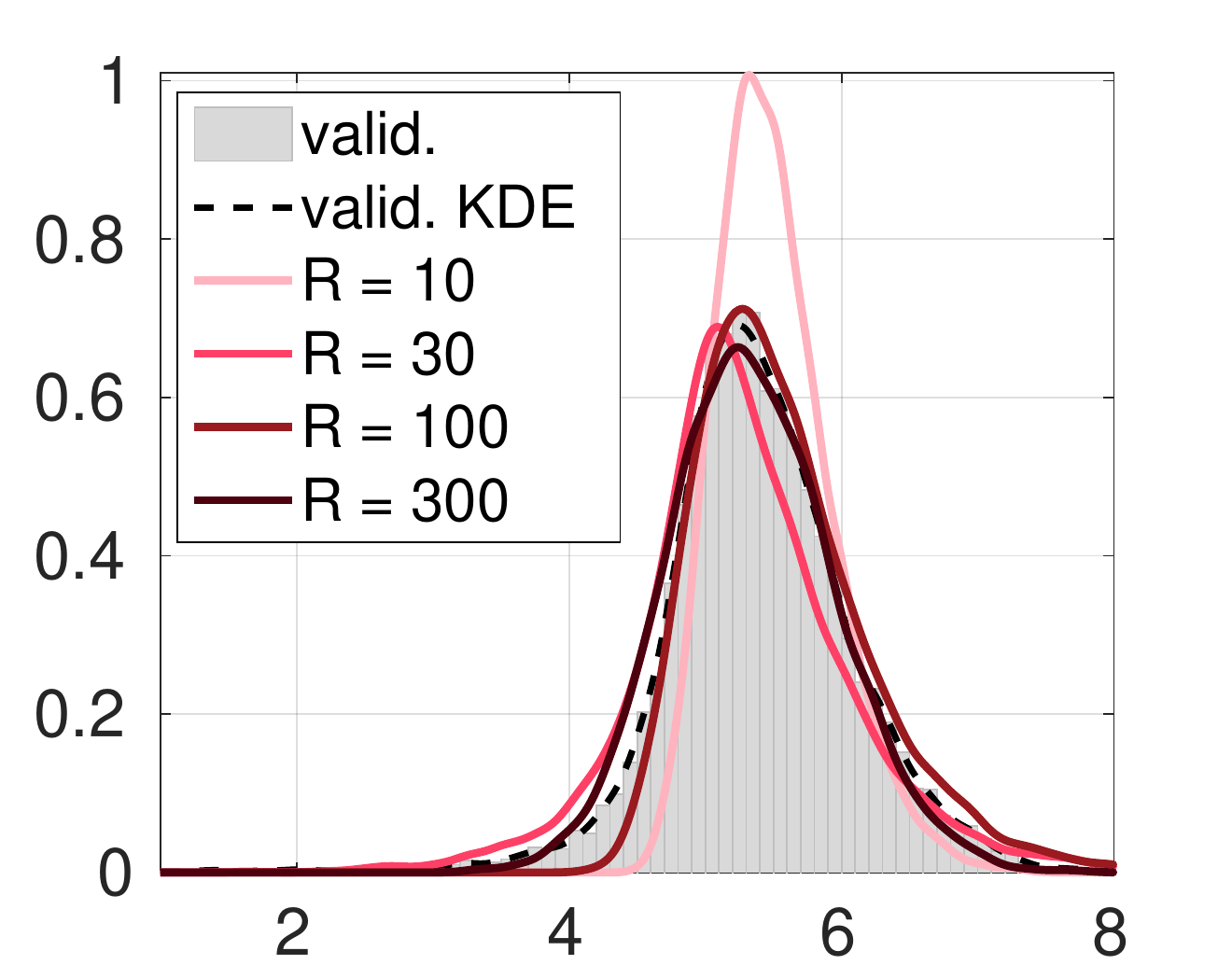}
		\label{fig:ishigami_conv_marg_vis}
	}
	\hfill
	\subfloat[Convergence of $\epsilon_\text{cov}$]
	{\includegraphics[width=.7\textwidth, height=.18\textheight, keepaspectratio]{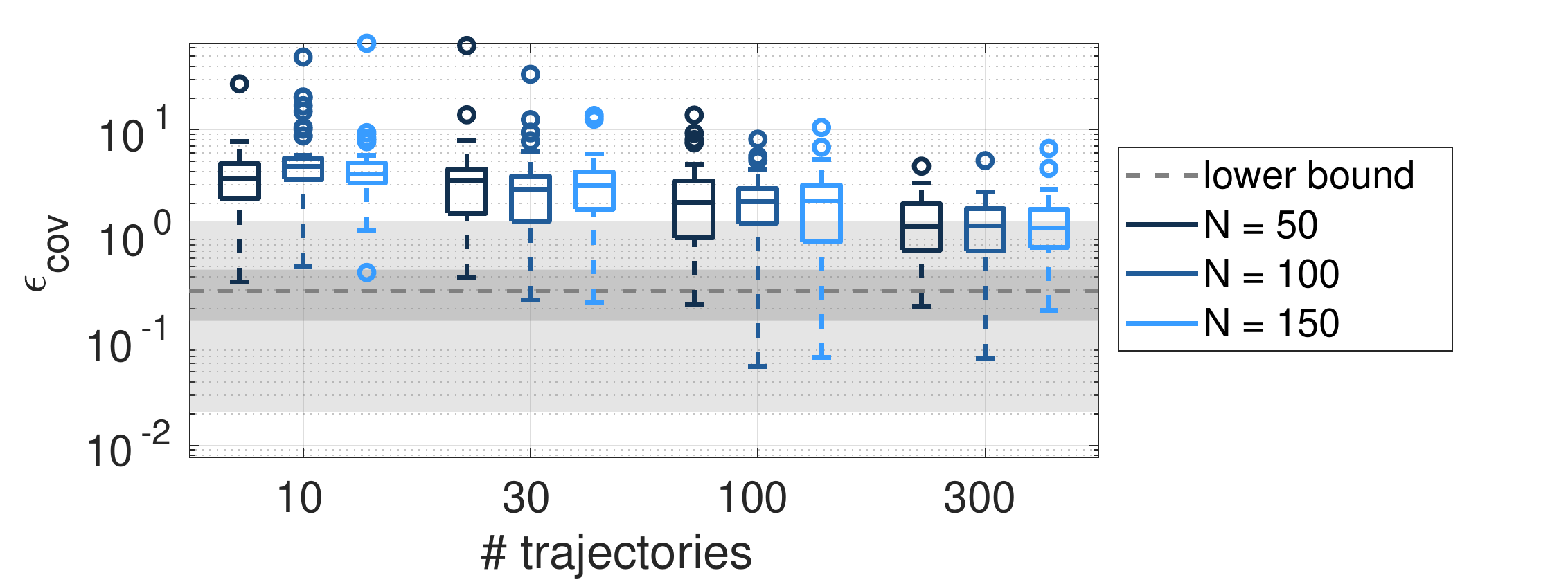}
		\label{fig:ishigami_convergence_covariance}}
	\caption{
		Convergence of $\epsilon_\text{marg}$ and $\epsilon_\text{cov}$ (\cref{eq:error_covariance,eq:wasserstein_avno}) for increasing number of available trajectories and parameter locations. Results for the stochastic emulator with parametric inference (choice~\ref{step:KLinf_paramEmu} of our algorithm in \cref{sec:stochemu_algorithm}) and 50 replications. The errors are computed based on a validation set of size $\Nparams_\text{val} = 1,000$, $\Nlatent_\text{val} = 10,000$.
		The gray areas and the dashed line represent quantiles and the median of a lower bound estimate for the respective error measure computed from $100\times 2$ independent MC samples of size $\Nlatent_\text{val} = 10,000$ generated by the true stochastic simulator. 
		The colored points in \cref{fig:ishigami_convergence_marginals} correspond to the results for a single replication and validation point as shown in \cref{fig:ishigami_conv_marg_vis} and help assess the meaning of the numerical error measures.
	}
	\label{fig:ishigami_convergence}
\end{figure}

In \cref{fig:ishigami_convergence_marginals}, we display the global convergence of marginal predictions for the parametric stochastic emulator in terms of $\epsilon_\text{marg}$ defined in \cref{eq:wasserstein_avno}.
Each boxplot represents one experiment (i.e., a specific number of experimental design points and number of trajectories), repeated independently 50 times.
The value of the averaged and normalized Wasserstein distance $\epsilon_\text{marg}$ is, by itself, difficult to interpret. To aid the interpretation and give an idea of the quality of the approximation, we add two auxiliary quantities to the plot:
\begin{itemize}
	\item The averaged and normalized Wasserstein distance is computed based on samples. As a lower bound, we 
	independently sample $100\times2$ validation sets (each consisting of $\Nlatent_\text{val} = 10,000$ trajectories evaluated at $\Nparam_\text{val} = 1,000$ points in the input domain; each pair of validation sets shares the same points in the input domain).
	We then compute the error in \cref{eq:wasserstein_avno} for each of the 100 pairs. The median and quantiles ($0.25$--$0.75$ and $0.05$--$0.95$) of this value are displayed in \cref{fig:ishigami_convergence_marginals} in gray, indicating the best possible error that can be achieved due to the natural variability of the sample estimates.
	\item A priori, it is unclear which value of the (averaged) normalized Wasserstein distance corresponds to predicted marginals that are visually close to the true marginals.
	To have some concrete examples on what a specific value of the normalized Wasserstein distance means,
	we consider the marginals predicted at one chosen validation point, shown in \cref{fig:ishigami_conv_marg_vis}. 
	We add the corresponding value of the normalized Wasserstein distance between simulator and emulator prediction as a small colored circle to the plot in \cref{fig:ishigami_convergence_marginals}.
\end{itemize}
We observe that the quality of the marginal estimates improves as we increase the size of the input parameter sample, which is expected since the PCE approximation of the trajectories becomes better with increasing experimental design size. $\Nparams = 50$ points are clearly too few to achieve a good estimate, whereas $\Nparams=100$ and $\Nparams = 150$ show convergence with the number of trajectories, indicating that the error from statistical inference is the dominating one. 
While the convergence of the marginals with the number of trajectories looks slow, the improvement of the marginal shapes is actually significant (compare the values with \cref{fig:ishigami_conv_marg_vis}).

In addition to marginal predictions, our stochastic emulator can also emulate the covariance function, using \cref{eq:covfct_series}. Since this equation relies only on the KL eigenfunctions, not on the KL-RV, the choice of inference method in Step~\ref{step:KLRVinference} of our algorithm in \cref{sec:stochemu_algorithm} does not affect these results.
In \cref{fig:ishigami_convergence_covariance} we display the convergence of $\epsilon_\text{cov}$ from \cref{eq:error_covariance}.
We observe that the error decreases with increasing numbers of trajectories. For the largest numbers of trajectories and experimental design points, the error is already in the range of the rough lower bound on achievable accuracy (obtained as described above by empirical sampling of the true model).
Interestingly, unlike the marginal error in \cref{fig:ishigami_convergence_marginals}, an increasing number $\Nparam$ of input parameter samples does not lead to a smaller covariance error. This indicates that the covariance estimate is less sensitive to the quality of the trajectory approximation, while the inference of the distribution of the KL-RV is more sensitive to it.

\begin{figure}[htb]
	\centering 
	{\includegraphics[width=.8\textwidth, height=.18\textheight, keepaspectratio, trim=0cm 0cm 1.2cm .4cm, clip]
		{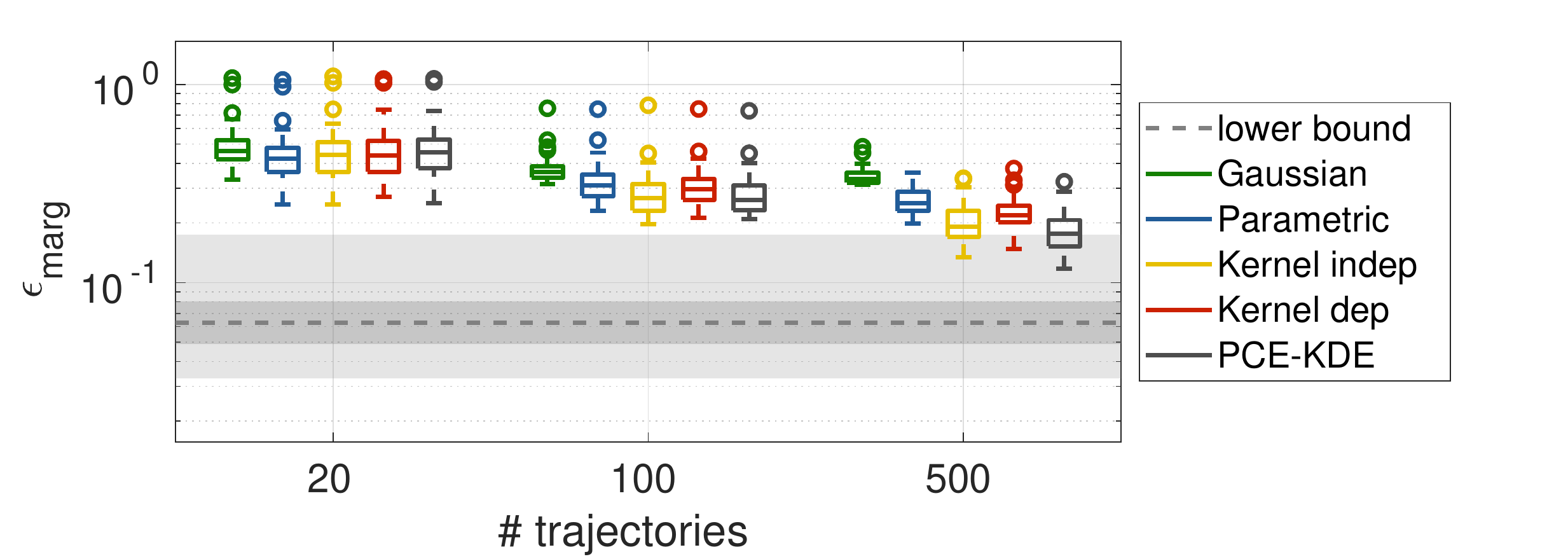}}
	\caption{Convergence of average normalized Wasserstein distance. Comparison of the four different methods for inferring the joint distribution of KL-RV described in Step~\ref{step:KLRVinference} 
		with the method PCE-KDE described in \cref{remark:PCE-KDE}. 
		$\Nparams = 100$ and $p = 14$. Errors are computed based on $\Nparams_\text{val} = 1,000$, $\Nlatent_\text{val} = 10,000$, for 50 replications.
		The gray areas and dashed line have the same meaning as in \cref{fig:ishigami_convergence}.
	}
	\label{fig:ishigami_convergence_othermethods}
\end{figure}

So far, we showed results for the stochastic emulator with parametric inference only (Option \ref{step:KLinf_paramEmu}). 
Now, in \cref{fig:ishigami_convergence_othermethods} we compare the four inference options described in Step~\ref{step:KLRVinference} of the algorithm in \cref{sec:stochemu_algorithm} with the results of a fifth method described in \cref{remark:PCE-KDE}, which we call here PCE-KDE.
We use the experimental design size $\Nparams = 100$, which yields PCE approximations with relative validation error of $10^{-5}$.
Due to this close fit, the PCE-KDE estimate can be considered as near-optimal estimate given the available data.

The error $\epsilon_\text{marg}$ is again computed on a validation set consisting of $\Nlatent_\text{val} = 10,000$ trajectories evaluated at $\Nparam_\text{val} = 1,000$ points in the input space.

We observe that for 20 trajectories, the five methods show almost identical performance. This suggests that 20 trajectories are not yet enough to infer a distribution that is able to generalize to unseen data, so that any marginal distribution with mean zero and unit standard deviation provides a reasonable approximation.
Comparing with the results for PCE-KDE, we see that our emulator is similarly accurate in prediction at an unseen point as a kernel density estimate using the training set of highly accurate PCE trajectories. This suggests that we do not lose much accuracy by applying our KLE approach on top of the PCE approximation, which can be seen as a form of dimension reduction in the stochastic space.

For the larger number of trajectories, $\Nlatent = 100$ and $\Nlatent = 500$, we do observe a  difference between the performance of the different marginal inference methods: standard Gaussians perform worst, while kernel density estimation without copula performs best of all the inference methods considered. Kernel density estimation with independence assumption performs almost on par with the PCE-KDE estimate.  
This suggests that the KL-RV are close to independent in this case, and that fitting a vine copula (using the available pair copulas) does not improve the overall inference, at least in the considered cases of limited data.
It also demonstrates that the true KL-RV distribution is not well approximated by independent standard Gaussians nor by other currently available parametric families, but that it can be approximated well by the more flexible kernel density estimation.
Parametric inference, offering a variety of standard marginal shapes, performs slightly better than Gaussian random variables.

\FloatBarrier

\subsection{Borehole function with latent variables}
\label{sec:borehole}
As a second example, we consider the well-known borehole function, which computes the water flow between two aquifers that are connected by a borehole \citep{Harper1983}.
It depends on eight parameters and is defined by
\begin{equation}
B(R_w, H_u, K_w, R, T_u, T_l, H_l, L) = \frac{2\pi T_u (H_u - H_l)}{\ln\left(R / R_w\right) \left( 1 + \frac{2 L T_u}{\ln\left(R / R_w\right) R_w^2 K_w} + \frac{T_u}{T_l}\right)}.
\label{eq:borehole_original}
\end{equation}
Its input random variables and their distributions are provided in Table \ref{table:borehole}.

We consider five parameters ($\ve \Xi = (R, T_u, T_l, H_l, L)$) of the borehole function to be latent, resulting in the three-dimensional stochastic simulator $\tilde B(r_w, h_u, k_w) = B(r_w, h_u, k_w; \ve \Xi)$.

For the three-dimensional input space, we use $\Nparam = \{20, 30, 60\}$ input samples and a maximal PCE degree of $p = 6$. 
The accuracy of the borehole approximation in terms of relative mean-squared validation error is in the order of  $10^{-3} / 10^{-7} / 10^{-10}$ for the different experimental design sizes.
The number of trajectories is in the range $\Nlatent = \{10,30,100,300\}$.

This results in typically $\Nkl = 2$ KL modes for an explained variance threshold of $99.9\%$. 
The eigenvalues of the KLE are approximately $\lambda_1 \approx 170$ and $\lambda_2 \approx 0.5$. The first mode alone covers more than $99.5\%$ of the total variance,
even though five independent parameters are used as latent variables.
Two of these have a significant total Sobol' index, and the sum of the first-order indices of the latent group is 19\%.
We will investigate this phenomenon in more detail in \cref{sec:numbermodes} below.

\begin{table}[htb]
	\small
	\caption{Borehole function: Input random variables and their distributions. For the borehole stochastic simulator with hidden variables, five of the eight variables (marked by italic letters) are considered latent.}
	\label{table:borehole}
	\centering
	\begin{tabular}{l l l l}
		\hline
		Variable & Distribution & Description & Total Sobol' index \\
		\hline
		$R_w$ & $\cn(0.10, 0.0161812)$ & borehole radius & 6.94e-01 \\
		$H_u$ & $\cu([990, 1110])$ & potentiometric head of upper aquifer & 1.06e-01 \\
		$K_w$ & $\cu([9855, 12045])$ & borehole hydraulic conductivity & 2.51e-02 \\
		$R$ & $ \text{Lognormal}([7.71, 1.0056])$ & \textit{radius of influence} & \it 2.77e-06 \\
		$T_u$ & $\cu([63070, 115600])$ & \textit{transmissivity of upper aquifer} & \it 2.10e-08 \\
		$T_l$ & $\cu([63.1, 116])$ & \textit{transmissivity of lower aquifer} & \it 8.23e-06 \\
		$H_l$ & $\cu([700, 820])$ &  \textit{potentiometric head of lower aquifer}& \it 1.06e-01 \\
		$L$ & $\cu([1120, 1680])$ & \textit{borehole length}& \it 1.03e-01 \\
		\hline
	\end{tabular}
\end{table}

As before, we analyze the global convergence of the marginal and covariance approximation for increasing numbers of input samples and trajectories, and we compare different methods for inferring the distribution of the KL-RV as described in Step~\ref{step:KLRVinference} of our algorithm (\cref{sec:stochemu_algorithm}). 
For the detailed explanation of the error measures, the setup of the convergence study, and the interpretation of the plots, see \cref{sec:numerical_experiments,sec:ishigami}.

\begin{figure}[htb]
	\centering
	\subfloat[Convergence of $\epsilon_\text{marg}$]
	{\includegraphics[width=.7\textwidth, trim=.2cm 0 1.3cm .1cm, clip, height=.19\textheight, keepaspectratio]
		{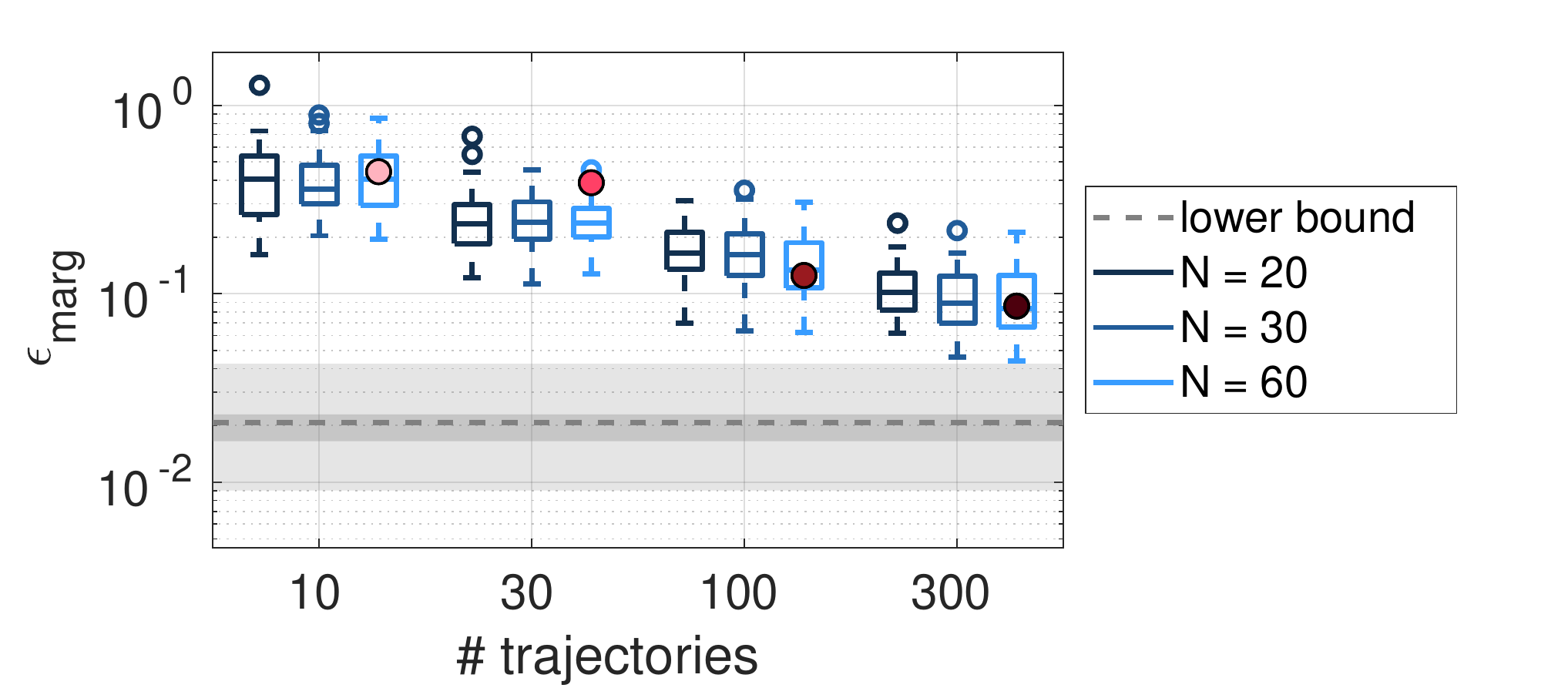}
		\label{fig:borehole_convergence_marginals}}
	\hfill	
	\subfloat[Marginal of $Y_{\ve x_\text{val}^{(1)}}$]
	{\includegraphics[width=.35\textwidth]
		{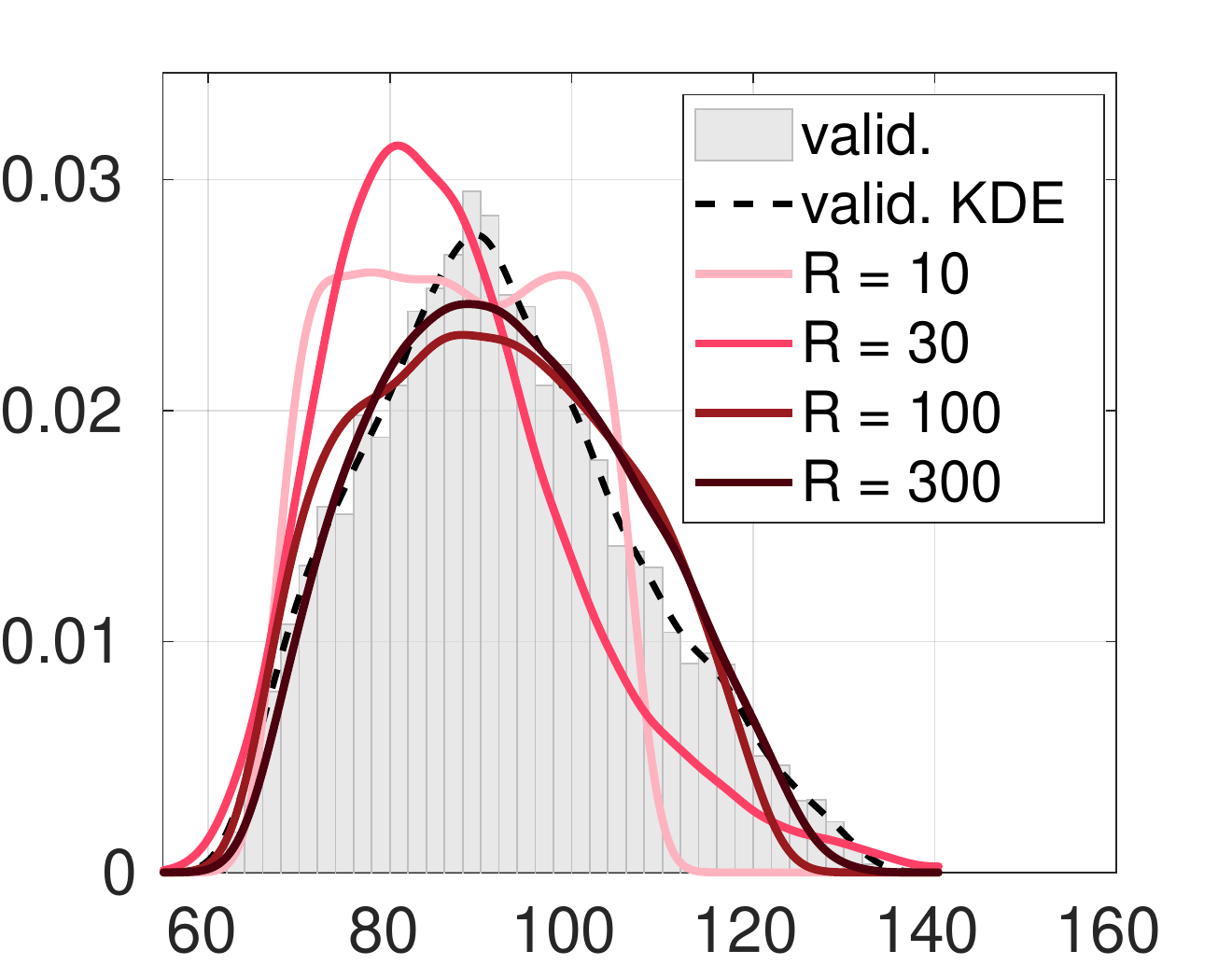}
		\label{fig:borehole_conv_marg_vis}
	}
	\\
	\subfloat[Convergence of $\epsilon_\text{cov}$]
	{\includegraphics[width=.7\textwidth, height=.18\textheight, keepaspectratio]{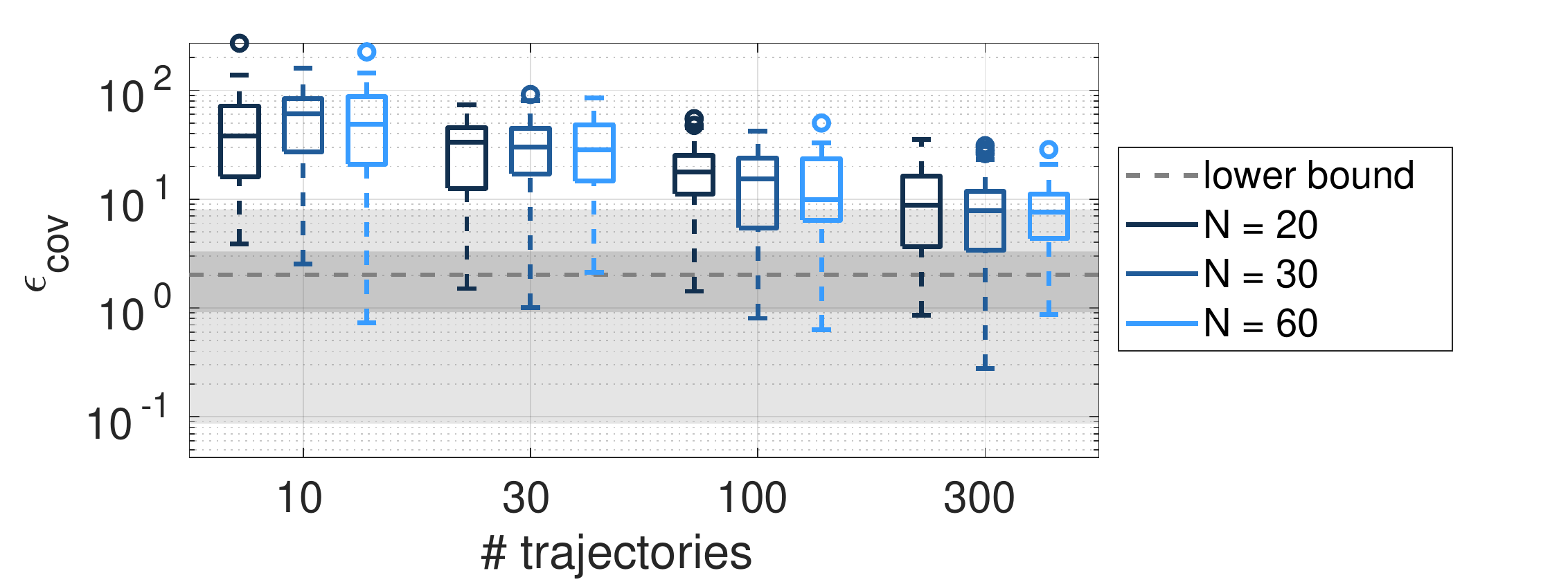}
		\label{fig:borehole_convergence_covariance}}
	\caption{Convergence of the $\epsilon_\text{marg}$ and $\epsilon_\text{cov}$ (\cref{eq:wasserstein_avno,eq:error_covariance}) for an increasing number of available trajectories and parameter locations. Results for the stochastic emulator with parametric inference (Option~\ref{step:KLinf_paramEmu}) and 50 replications. The errors are computed based on a validation set of size $\Nparams_\text{val} = 1,000$, $\Nlatent_\text{val} = 10,000$.
		The gray areas and the dashed line represent quantiles and the median of a lower bound estimate for the respective error measure computed from $100 \times 2$ independent MC samples of size $\Nlatent_\text{val} = 10,000$ generated by the true stochastic simulator. 
		The colored points in \cref{fig:borehole_convergence_marginals} correspond to the results for a single replication and validation point as shown in \cref{fig:borehole_conv_marg_vis} and help assess the meaning of the numerical error measures.
	}
	\label{fig:borehole_convergence}
\end{figure}

In \cref{fig:borehole_convergence_marginals}, we see that our method converges in both global error metrics ($\epsilon_\text{marg}$ and  $\epsilon_\text{cov}$) towards the rough empirical lower bound indicated by the gray area and dashed line. For $\epsilon_\text{marg}$, the difference between the results for the three experimental design sizes $\Nparams = 20, 30$, and $60$ is small. This indicates that at least for this model, a validation mean-squared error smaller than {$\co(10^{-3})$} does not lead to significantly more accurate results, and that below this accuracy the error is dominated by the inference error.

The convergence of the emulated covariance function is displayed in \cref{fig:borehole_convergence_covariance}.
As expected, the global error become smaller with an increasing number of trajectories, and approaches the lower bound representing the variability due to the error being computed from samples.
Again, the difference in the results for the three experimental design sizes $\Nparam = 20, 30, 60$ is small. 
Since the first mode accounts for more than $99.5\%$ of the explained variance, the first KLE eigenfunction has the dominating influence on the covariance estimation (\cref{eq:covfct_series}). The results indicate that the first eigenfunction and its eigenvalue are estimated accurately already for the smallest experimental design sizes.

\begin{figure}[htb]
	\centering
	{\includegraphics[width=.8\textwidth, height=.18\textheight, keepaspectratio, trim=0cm 0cm 1.2cm .4cm, clip]
		{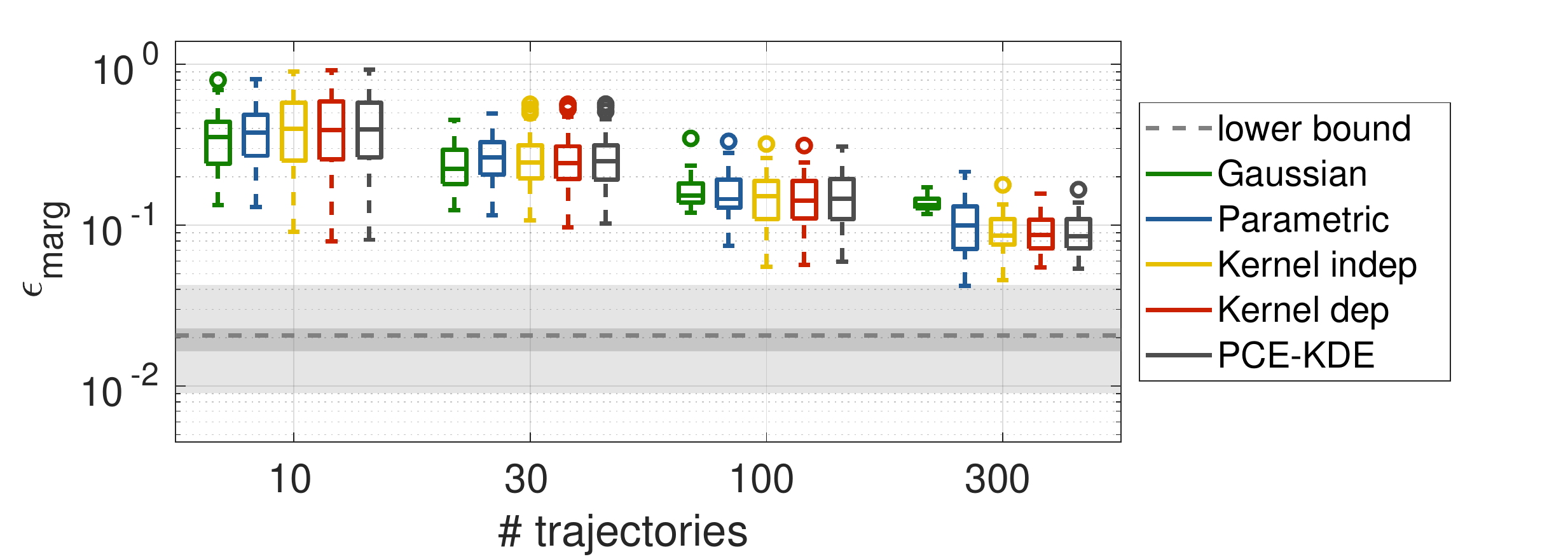}}
	\caption{Convergence of $\epsilon_\text{marg}$. Comparison of the four different methods for inferring the joint distribution of KL-RV described in Step~\ref{step:KLRVinference} with PCE-KDE described in \cref{remark:PCE-KDE}.
		$\Nparams = 60$ and $p = 6$. Errors are computed based on $\Nparams_\text{val} = 1,000$, $\Nlatent_\text{val} = 10,000$, and 50 replications.
		The gray areas and dashed line have the same meaning as in \cref{fig:borehole_convergence}.
	}
	\label{fig:borehole_convergence_othermethods}
\end{figure}

The comparison of the different inference methods for the distribution of the KL-RV (Step~\ref{step:KLRVinference} of the algorithm) is displayed in \cref{fig:borehole_convergence_othermethods}. 
Similarly as for the Ishigami function, we observe that for a small number of trajectories ($\Nlatent = 10$ and $30$) the four inference methods and PCE-KDE show almost the same performance. 
Modeling the KL-RV with standard Gaussian distributions seems to offer a slight advantage (resulting in a slightly smaller median error and smaller variability) 
for small numbers of trajectories, probably because they make the strongest assumptions on the distribution shape, which is advantageous for generalizability in the case of small data. 
As the number of trajectories grows, a similar pattern as in \cref{sec:ishigami} emerges: standard Gaussian inference shows the worst performance, followed by parametric inference. Inference with kernel density estimation (dependent and independent) shows the best performance, on par with the PCE-KDE estimate, which (due to the high accuracy of the PCE approximations for $\Nparams=60$) represents the near-optimal estimate given the available training data.

Interestingly, there is no significant difference between the performance of KDE with and without the independence assumption. Here the magnitude of the eigenvalues might offer an explanation: with more than two orders of magnitude difference between $\lambda_1$ and $\lambda_2$, the dependence between the two random variables $\xi_1$ and $\xi_2$ does not influence the resulting predictions as much as the correct identification of the marginal shape of the first KL-RV $\xi_1$.

\FloatBarrier

\subsection{Heston stochastic volatility model for a stock price}
\label{sec:heston}
As a third example, we consider the Heston stochastic volatility model, which describes a stock price $Y_t$ \citep{Heston1993} with its volatility $\nu_t$ modeled as stochastic process:
\begin{align}
\di{U_t} &= \mu U_t \di{t} + \sqrt{\nu_t}U_t \di{W_t^{(1)}}, 
\label{eq:Heston1}\\
\di{\nu_t} &= \kappa (\theta - \nu_t) \di{t} + \sigma \sqrt{\nu_t} \di{W_t^{(2)}}
\label{eq:Heston2}
\end{align}
with two Wiener processes $W_t^{(1)}$ and $W_t^{(2)}$ with correlation coefficient $\rho$.
This model has six uniformly distributed parameters $\ve X = (\mu, \kappa, \theta, \sigma, \rho, \nu_0)$ detailed in \cref{tab:heston_params}, the bounds of which are calibrated from real data as described in \citet{ZhuRESS2021}. 
The quantity of interest is 
\begin{equation}
Y_{\ve x} = U_1(\ve X = \ve x),
\end{equation}
i.e., the stock price after 1 year. As proposed by \citet{ZhuRESS2021}, we set $U_0 = 1$ and use the Euler-Maruyama method to integrate the system of stochastic differential equations (SDEs) and replace $\nu_t$ by $\max(\nu_t, 0)$ to avoid negative values of $\nu_t$.
This model is a stochastic simulator due to the stochasticity induced by the two Wiener processes $W_t^{(1)}$ and $W_t^{(2)}$ driving the SDEs. A trajectory in the parameter space $\cd$ is obtained by fixing the realizations of these processes and evaluating \cref{eq:Heston1,eq:Heston2} for $\ve x \in \cd$.

For the six-dimensional input space, we use $\Nparam = \{50, 100, 150\}$ input samples and a maximal PCE degree of $p = 7$. The accuracy of the approximation in terms of relative mean-squared validation error is ca.\ $\co(0.03) / \co(0.02) / \co(0.006)$ for the different experimental design sizes. 
This means that the Heston model is not particularly well approximated by PCE, even for rather large experimental designs.
We use a number of trajectories in the range $\Nlatent = \{10, 30, 100, 300\}$.

This results in typically $\Nkl = 4$ to $6$ KL modes for an explained variance threshold of $99.9\%$. The first eigenvalue is $\lambda_1 \approx 0.05$ and usually covers more than 97\% of the variance.

\begin{table}[htb]
	\centering
	\small
	\caption{Parameters and their distributions for the Heston SDE model.}
	\label{tab:heston_params}
	\begin{tabular}{lll}
		\hline
		Variable & Distribution & Description \\
		\hline
		$\mu$ & $\cu([0, 0.1])$ & Expected return rate \\
		$\kappa$ & $\cu([0.3, 2])$ & Mean reversion speed of the volatility \\
		$\theta$ & $\cu([0.02, 0.07])$ & Long term mean of the volatility \\
		$\sigma$ & $\cu([0.2, 0.4])$ & Volatility of the volatility \\
		$\rho$ & $\cu([-1, -0.5])$ & Correlation coefficient between $\di{W_t^{(1)}}$ and $\di{W_t^{(2)}}$ \\
		$\nu_0$ & $\cu([0.02, 0.07])$ & Initial volatility\\
		\hline
	\end{tabular}
\end{table}

\begin{figure}[htb]
	\centering
	\subfloat[Convergence of $\epsilon_\text{marg}$]
	{\includegraphics[width=.7\textwidth, trim=0cm 0cm 1cm 0cm, clip, height=.19\textheight, keepaspectratio]
		{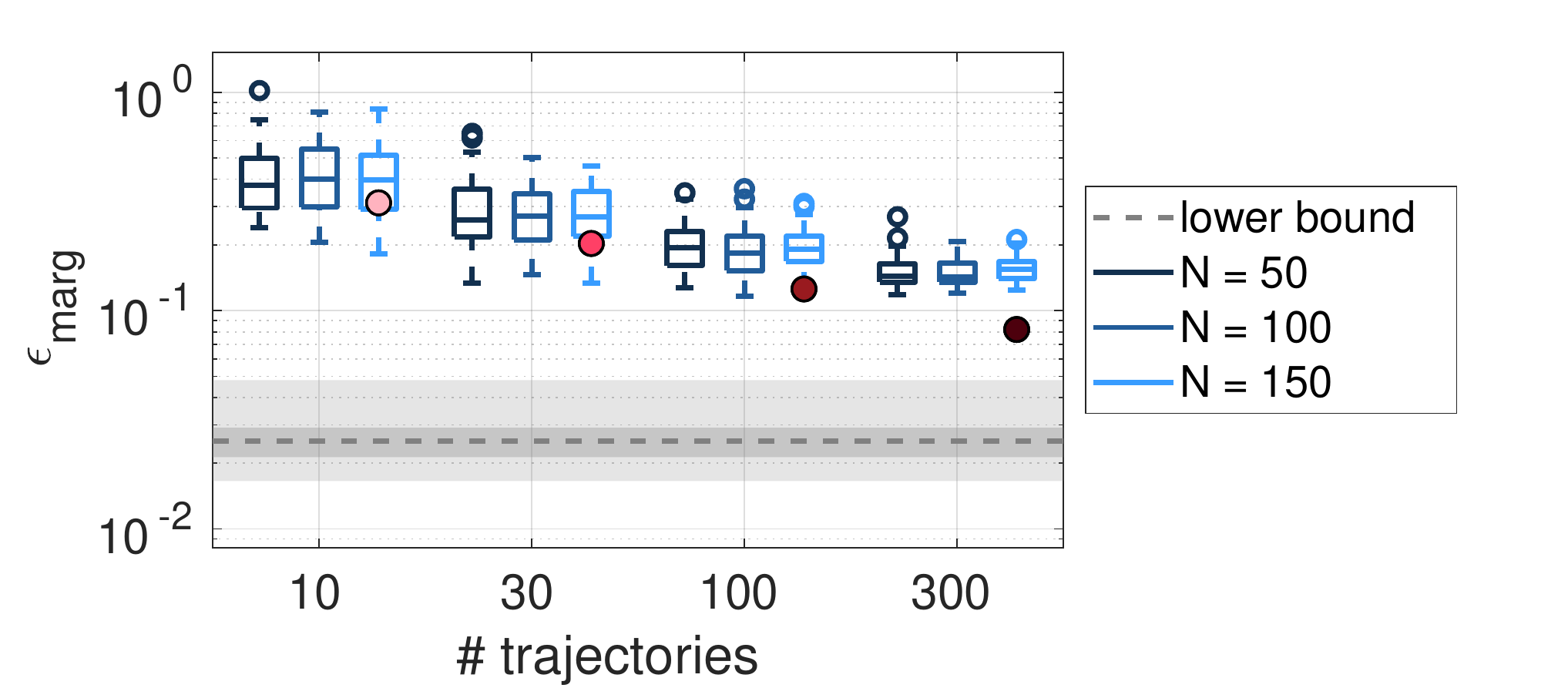}\label{fig:heston_convergence_marginals}}
	\hfill
	\subfloat[Marginal of $Y_{\ve x_\text{val}^{(1)}}$]
	{\includegraphics[width=.35\textwidth]
		{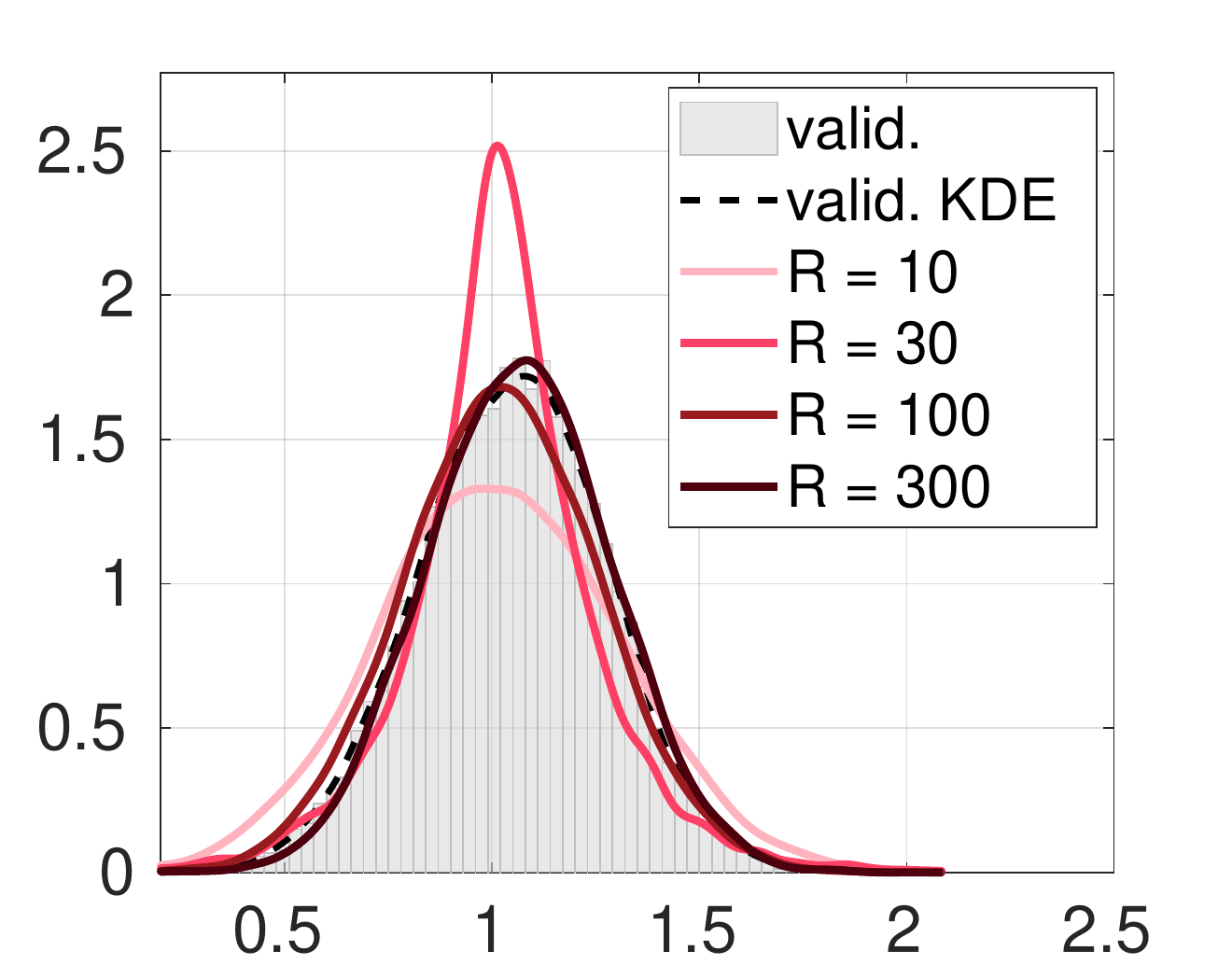}\label{fig:heston_conv_marg_vis}}
	\\
	\subfloat[Convergence of $\epsilon_\text{cov}$]
	{\includegraphics[width=7\textwidth, height=.18\textheight, keepaspectratio]
		{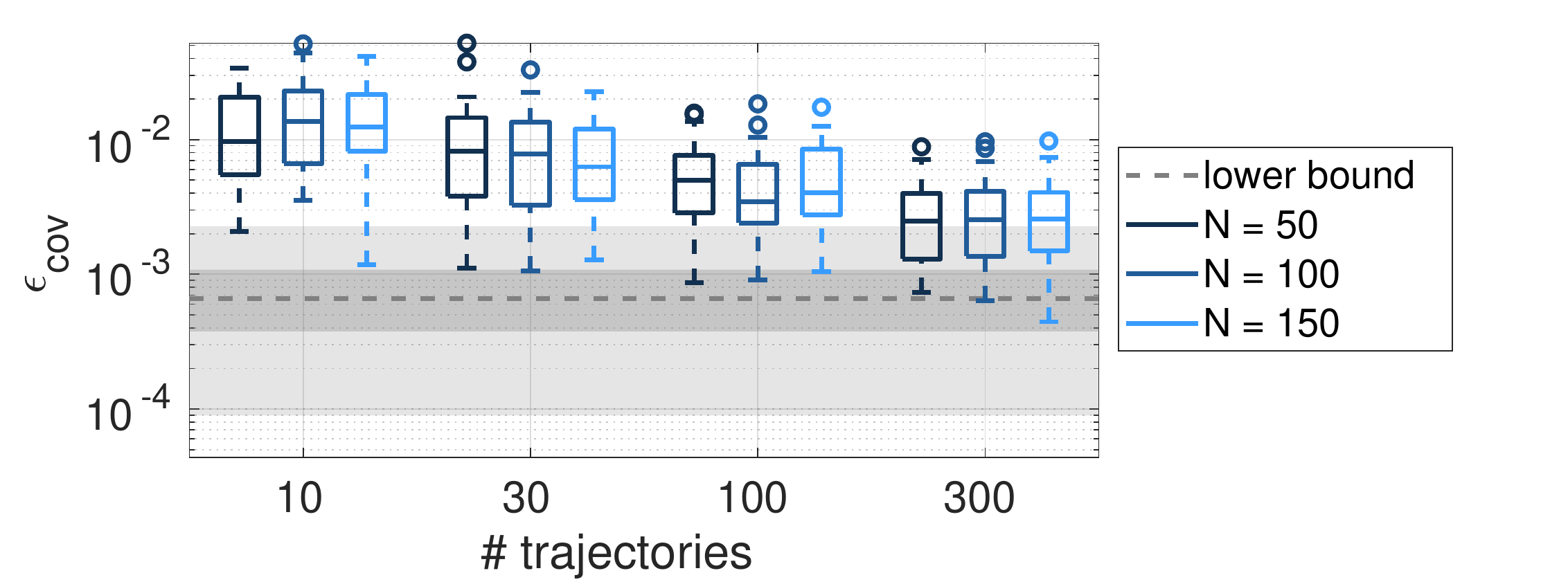}\label{fig:heston_convergence_covariance}}
	\caption{
		Convergence of $\epsilon_\text{marg}$ and $\epsilon_\text{cov}$ (\cref{eq:wasserstein_avno,eq:error_covariance}) for an increasing number of available trajectories and parameter locations. Results for the stochastic emulator with parametric inference (choice~\ref{step:KLinf_paramEmu}) and 50 replications. The errors are computed based on a validation set of size $\Nparams_\text{val} = 1,000$, $\Nlatent_\text{val} = 10,000$.
		The gray areas and the dashed line represent quantiles and the median of a lower bound estimate for the respective error measure computed from $100 \times 2$ independent MC samples of size $\Nlatent_\text{val} = 10,000$ generated by the true stochastic simulator. 
		The colored points in \cref{fig:heston_convergence_marginals} correspond to the results for a single replication and validation point as shown in \cref{fig:heston_conv_marg_vis} and help assess the meaning of the numerical error measures.
	}
	\label{fig:heston_convergence}
\end{figure}

Again, we analyze the global convergence of the marginal and covariance approximation in the same way as in the preceding sections.
The marginal approximations of the parametric stochastic emulator converge with increasing experimental design size and number of trajectories, but slowly, as displayed in \cref{fig:heston_convergence_marginals}.
There is no significant difference between the three experimental design sizes $\Nparam = 50, 100, 150$.
This indicates that the improvement due to a better PCE approximation for an increasing number of experimental design points is overshadowed by the inaccuracy due to the inference of the KL-RV.
This, in turn, could be because the PCE approximation is not yet sufficiently accurate (note that the relative validation error is in the order of $10^{-2}$ for all ED sizes.)
Even for the largest number of trajectories, the averaged and normalized Wasserstein distance between the responses of the true model and the emulator is still much larger than the variability resulting from sampling the true model, which is illustrated by the gray areas and the dashed line in \cref{fig:heston_convergence_marginals} (quantiles and median, respectively).
Comparing the boxplots to the colored points corresponding to the marginal estimates illustrated in \cref{fig:heston_conv_marg_vis}, we observe that the marginal shape of the stochastic response for one validation point $\ve x_\text{val}^{(1)}$ is already captured quite well for 100-300 trajectories (with the value for $\Nlatent=300$ being a bit of an outlier).

The convergence of the covariance function is shown in \cref{fig:heston_convergence_covariance}. 
As expected, the covariance estimate becomes better with increasing number of trajectories, even approaching the lower bound obtained by resampling the original model. However, we observe again that an increasing number of experimental design points does not influence the estimate much, which indicates that the covariance estimate is quite robust against the trajectory approximation quality. 
Since the first mode is also dominant for this example (accounting for ca.\ $97 \%$ of the explained variance), it indicates that the first eigenfunction is well estimated already for small experimental design sizes.

\begin{figure}[htb]
	\centering
	{\includegraphics[width=.8\textwidth, trim=0cm 0cm 1.4cm .5cm, clip, height=.18\textheight, keepaspectratio]
		{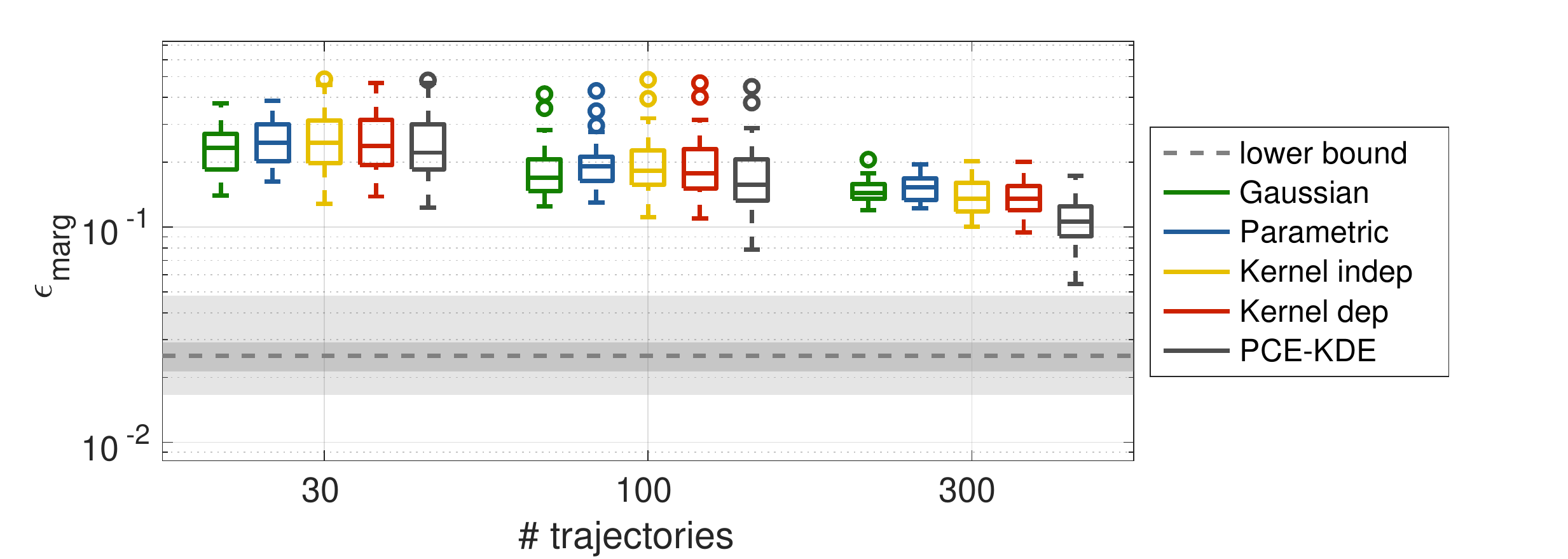}}
	\caption{Convergence of marginals (Wasserstein distance). Comparison of the four different methods for inferring the joint distribution of KL-RV described in Step~\ref{step:KLRVinference} with PCE-KDE described in \cref{remark:PCE-KDE}.
		$\Nparams = 150$ and $p = 5$. Errors are computed based on $\Nparams_\text{val} = 1,000$, $\Nlatent_\text{val} = 10,000$, and 50 replications.
		The gray areas and dashed line have the same meaning as in \cref{fig:heston_convergence}.
	}
	\label{fig:heston_convergence_othermethods}
\end{figure}

\cref{fig:heston_convergence_othermethods} shows the comparison between the different methods for KL-RV inference (Step~\ref{step:KLRVinference} of the algorithm in \cref{sec:our_approach}) as described in \cref{sec:ishigami_convergence}. 
All four methods perform comparably.
The independent standard Gaussian approximation performs slightly better than the other methods in the case of few trajectories and slightly worse for the case of many trajectories, which is consistent with the previous cases.
Again, KDE with and without dependence shows almost identical performance, indicating that the either the true copula is the independence copula, or that the existing parametric copulas are not suitable for capturing the dependence structure.
While the inference methods show a similar performance to PCE-KDE for smaller numbers of trajectories, PCE-KDE finds a better marginal approximation when $\Nlatent = 300$ trajectories are available. This indicates that some information is lost in the KLE procedure of \cref{sec:our_approach}.

\FloatBarrier

\section{Considerations on the number of modes}
\label{sec:numbermodes}

In this section we investigate how many modes $K$ we can expect in the stochastic emulator of \cref{eq:stochastic_emulator}.
We consider here a certain class of stochastic simulators that arise from a deterministic model $\ve z \mapsto \cm(\ve z)$ by considering some of its variables as \emph{hidden} (or \emph{latent}). In other words, the stochastic simulator is $\cm(\ve X, \ve \Xi)$, where $\ve X$ are the explicit input variables, and $\ve \Xi$ the latent variables inducing the stochasticity (the random events, see \cref{sec:random_fields}). Assume that all components of $\ve Z = (\ve X, \ve \Xi)$ are independent, and denote by $f_k$ the marginal distribution of $Z_k$.

Assume further that the deterministic simulator $\cm$ has finite variance under $f_{\ve Z}$. Then it can be decomposed into the Hoeffding-Sobol' decomposition (a.k.a.\ ANOVA decomposition, analysis of variance) \citep{Hoeffding1948,Sobol1995} as
\begin{equation}
\cm(\ve Z) = m_0 + \sum_{1 \leq i \leq d} m_i(Z_i) + \sum_{1 \leq i <j \leq d} m_{i,j}(Z_i, Z_j) 
+ \dots + m_{1, \dots, d}(Z_1, \dots, Z_d)
\label{eq:sobolhoeffding}
\end{equation}
where the terms satisfy $\int m_I(\ve Z_I) f_{k}(z_k) \di{z_k} = 0$ for all $k \in I \subset \{1 \enum d\}$.
$m_0$ is the mean of $\cm(\ve Z)$. The univariate terms $m_i$ are called \emph{main effects}, and the remaining summands are \emph{interaction terms} of increasing order.

Now we group the summands of \cref{eq:sobolhoeffding} according to whether they involve only input variables, only latent variables, or some variables from both groups. This results in the following decomposition:
\begin{equation}
\cm(\ve X, \ve \Xi) = m_0 + \cm_1(\ve X) + \cm_2(\ve \Xi) + \cm_{12}(\ve X, \ve \Xi)
\label{eq:stochsim_decomp}
\end{equation}
where, e.g., $\cm_1(\ve x) = \sum_{I: {\ve Z}_I \subset \ve X} m_I(\ve z_I)$.
The last summand of \cref{eq:stochsim_decomp} contains all interaction terms from \cref{eq:sobolhoeffding} that involve at least one input and at least one latent variable.

It is a rather common phenomenon in uncertainty quantification that engineering models actually have near-zero interaction terms. 
In that case, the model is essentially additive with respect to the two groups of variables $\ve X$ and $\ve \Xi$:
\begin{equation}
\label{eqn:additive}
\cm(\ve{X},\ve{\Xi}) \approx
m_0 + m_1(\ve X) + m_2(\ve \Xi).
\end{equation}
This means that any realization $\ve{\xi}$ of the unknown latent variables $\ve{\Xi}$ results in a constant shift of $\cm(\cdot, \ve \xi)$ regardless of the value of the input parameters $\ve{x}$, a behavior that can be accurately modeled by a single KL mode:
if equality holds in \cref{eqn:additive}, the mean function is $\mu(\ve x) = \Espe{\ve\Xi}{\cm(\ve{x},\ve{\Xi})} =  m_0 + m_1(\ve X)$, 
the covariance function is $c(\ve x, \ve x') = \Var{m_2(\ve \Xi)}$, and the only nonzero eigenvalue of \cref{eq:integral_EV_problem} is $\lambda_1 = \Var{m_2(\ve \Xi)}$ with eigenfunction $\phi_1(\ve x) = 1$.

We have observed this in the numerical examples in \cref{sec:numerical_experiments}, e.g., for the case of the borehole model with hidden variables. The deterministic borehole model defined in \cref{eq:borehole_original} has relatively low interaction: despite its nonlinearity, under the input uncertainties in Table~\ref{table:borehole} its first order Sobol' indices \citep{Sobol1993} sum up to $\sum_{i=1}^8 S^1_i \approx  96.7\%$. 
The interaction effect between the explicit and the latent group is around 2\%.
Therefore, only one mode is sufficient to model the stochastic simulator that results from treating several of the model's variables as latent.

\section{Discussion and conclusions}
\label{sec:conclusion}

We presented a spectral surrogate model for stochastic simulators, a special class of computational models whose response for a given input is a random variable. 
Our method relies on several advanced techniques for modeling uncertainties, such as polynomial chaos expansion (PCE), Karhunen-Lo\`eve expansion (KLE), and statistical inference of multivariate distributions.
The resulting surrogate model is not only able to emulate the marginal distributions and the covariance structure, but it can also generate new trajectories.

The form of our surrogate model provides insight into the model structure. 
The number of expansion modes indicates how high-dimensional the underlying stochasticity is. 
The eigenfunctions of the KLE, which are polynomials, give information about how the input parameters influence the stochastic simulator output.
Even though our numerical examples were chosen to represent a range of cases of increasing complexity, we found that one mode was usually sufficient to explain more than $95\%$ of the variance of the stochastic simulator.
We were able to explain this phenomenon for the common case of stochastic simulators 
that arise from finite-dimensional deterministic models with independent inputs and finite variance by treating some of their input variables as latent.
Considering the Hoeffding-Sobol' decomposition of the underlying deterministic simulator, we found that if the interaction terms between the explicit and latent variables are negligible, one single KL mode will be sufficient to emulate the behavior of the stochastic simulator. 
Indeed, by experience, this is a common occurrence in applications of uncertainty quantification. Interactions are rarely dominant in engineering problems, and so one KL mode might suffice in many cases.

From our numerical experiments, we found that the Gaussian (or more generally, parametric) approximation of marginals of the KL-random variables (KL-RV) can be preferable if the number of available trajectories is very small. When more trajectories are available, the better choice is kernel density estimation.
Since the first mode was dominant in our numerical examples, the characterization of the dependence between KL-RV turned out not to be crucial, at least for the class of applications considered. 

Our numerical tests reveal that the emulator is able to capture the true model behavior reasonably well, as long as enough input samples and trajectories are used.  
Due to the sequential nature of our approach, it is important to use enough points in the experimental design: if the PCE approximation is not accurate enough, also the inferred distribution for the KL-RV will be inaccurate. 
Interestingly, we observed in all three examples that the covariance estimate was not heavily influenced by using a larger experimental design, even though the latter typically results in more accurate PCE approximations of the trajectories. This indicates that the number of trajectories is more important for the covariance approximation than the quality of the PCE approximation. Also, it seems that (since the first mode was dominant for all investigated models) the first KLE eigenfunction can be identified accurately already from a small experimental design.

Note that our surrogate relies on the assumption that the trajectories  are well approximated by sparse PCE, an assumption not fulfilled by some stochastic simulators, e.g., ones with discontinuous or non-differentiable trajectories in the input parameter space.
This could be circumvented by using another basis specially adapted to the purpose, such as wavelets.
Furthermore, by construction, our emulator is only suitable for stochastic simulators whose response is correlated throughout the input domain.
If there is little to no correlation between the responses at different points in the input space, KLE (which is ultimately a dimension reduction technique), would need infeasibly many modes to converge. 

Our methodology can be extended in several ways. 
The computation of the sparse PCE approximation of the trajectories could be done jointly for all trajectories, instead of fitting each trajectory separately and later discarding regressors. 
While in our study the dependence between KL-RV was not crucial for the accuracy of the emulator, an improved estimation of the dependence structure would be desirable if for future models more modes turn out to be important. 
Furthermore, an interesting question is under which circumstances one mode is enough to represent the underlying stochasticity of stochastic simulators, and how the methodology can be adapted to take advantage of this phenomenon.
The general idea of representing trajectories by their coefficients, and after dimension reduction modeling their joint distribution, can also be applied outside spectral expansions, e.g.\ for Bayesian neural networks, at the cost of losing some of the analytical properties following from orthogonality. 
Finally, in the spirit of \emph{common random numbers} \citep{Pearce2022}, the applicability of our stochastic emulator for purposes such as optimization should be explored.

\section*{Acknowledgments}
This paper is part of the project ``Surrogate Modeling for Stochastic Simulators (SAMOS)'' funded by the Swiss National Science Foundation (Grant \#200021\_175524), whose support is gratefully acknowledged.


\bibliography{stoch_sim_paper_neutral}

\appendix
\section{Analytical derivations for extended KLE on PCE trajectories}
\label{app:derivations_KLE_PCE}

The following is a detailed exposition of the analytical computations for extended KLE using the PCE basis in $\ivsp$. A less detailed derivation for $L^2(\cd)$ can be found in \citet[Section~8.4.2]{Ramsay2005}.

We show in \cref{app:derivations_KLE_PCE_eigenfunctions} that 
if the trajectories $\ve x \mapsto \cm(\ve x, \omega)$ are represented by PCE, and extended KLE is applied,
then the sample covariance function is a polynomial, the integral eigenvalue problem reduces to PCA in the expansion coefficients, and the eigenfunctions are polynomials. 
In \cref{app:derivations_KL-RV}, we show that the realizations of the random KLE coefficients can be determined analytically.

Let for $r = 1 \enum \Nlatent$ 
\begin{equation}
\tilde \cm^\text{PCE}_r(\ve x) = \sum_{\alp\in\ca} {\tilde \coeff}_\alp \psi_\alp(\ve x)
\end{equation}
be the centered PCE trajectory computed from discrete evaluations of the trajectory $\ct_r$ (\cref{eq:discrete_trajectories}).
The sample covariance function is a polynomial given by
\begin{equation}
\hat \covfct(\ve x, \ve x') = \frac{1}{\Nlatent-1} \sum_{r=1}^{\Nlatent} \tilde \cm^\text{PCE}_r(\ve x) \tilde \cm^\text{PCE}_r(\ve x').
\end{equation}

\subsection{Analytical solution of the extended KLE eigenvalue problem}
\label{app:derivations_KLE_PCE_eigenfunctions}

Computing an extended KLE in the function space $\ivsp$ corresponds to solving the following eigenproblem:
\begin{equation}
\left\langle \hat \covfct(\ve x, \cdot), \phi_i(\cdot) \right\rangle_{\!\ivsp} 
= \int_\cd \hat \covfct(\ve x,\ve x')\phi_i(\ve x') f_{\ve X}(\ve x') \ \di{\ve x'} 
= \lambda_i \phi_i(\ve x).
\label{eq:eKLE_eigenproblem}
\end{equation}
Since $\hat c$ is polynomial, also the eigenfunctions will be polynomials and can be represented in terms of the PCE basis as follows:
\begin{equation}
\phi_i(\ve x) = \sum_{\alp\in\ca} \eigv_\alp^{(i)} \psi_\alp(\ve x).
\end{equation}
Solving \cref{eq:eKLE_eigenproblem} reduces to finding $(\lambda_i, (\eigv_\alp^{(i)})_{\alp \in \ca})$ for $i = 1 \enum \Nkl$.

Dropping the $i$-subscript of the eigenfunction for convenience, we compute

\begin{align*}
\int_\cd \hat \covfct(\ve x,\ve x')\phi(\ve x') f_{\ve X}(\ve x') \ \di{\ve x'}
&= \int_\cd \frac{1}{\Nlatent - 1} \sum_{r=1}^{\Nlatent} \tilde{\cm}^\text{PCE}_r(\ve x) \tilde{\cm}^\text{PCE}_r(\ve x') \phi(\ve x') f_{\ve X}(\ve x')\ \di{\ve x'} \\
&= \frac{1}{\Nlatent - 1} \sum_{r=1}^{\Nlatent} \tilde{\cm}^\text{PCE}_r(\ve x) \int_\cd  \tilde{\cm}^\text{PCE}_r(\ve x') \phi(\ve x') f_{\ve X}(\ve x') \ \di{\ve x'} \\
&= \frac{1}{\Nlatent - 1} \sum_{r=1}^{\Nlatent} \tilde{\cm}^\text{PCE}_r(\ve x) \int_\cd \left( \sum_{\alp \in \ca} \tilde{\coeff}_\alp^r \psi_\alp(\ve x') \right) \left( \sum_{\ve\beta \in \ca} \eigv_{\ve \beta} \psi_{\ve \beta}(\ve x') \right) f_{\ve X}(\ve x') \ \di{\ve x'} \\
&= \frac{1}{\Nlatent - 1} \sum_{r=1}^{\Nlatent} \tilde{\cm}^\text{PCE}_r(\ve x) \left( \sum_{\alp \in \ca} \tilde{\coeff}_\alp^r \eigv_\alp \right) \qquad \text{(orthonormality of PCE basis)}\\
&= \frac{1}{\Nlatent - 1} \sum_{r=1}^{\Nlatent} \left(\sum_{\ve\beta \in \ca} \tilde{\coeff}_{\ve\beta}^r \psi_{\ve\beta}(\ve x) \right) \left( \sum_{\alp \in \ca} \tilde{\coeff}_\alp^r \eigv_\alp \right)\\
&= \sum_{\ve\beta \in \ca} \left( \frac{1}{\Nlatent - 1} \sum_{r=1}^{\Nlatent}  \tilde{\coeff}_{\ve\beta}^r  \left( \sum_{\alp \in \ca} \tilde{\coeff}_\alp^r \eigv_\alp \right)	 \right) \psi_{\ve\beta}(\ve x) .
\end{align*}
The eigenvalue problem reduces to
\begin{align*}
\sum_{\ve\beta \in \ca} \left( \frac{1}{\Nlatent - 1} \sum_{r=1}^{\Nlatent}  \tilde{\coeff}_{\ve\beta}^r
\left(
\smash[b]{\underbrace{ \sum_{\alp \in \ca} \tilde{\coeff}_\alp^r \eigv_\alp }_{= (\tilde{\ve \coeff}^r)^T \!\ve \eigv \, \in \, \Rr}} 
\right) 
\right) \psi_{\ve\beta}(x) 
\overset{!}{=} 
\lambda \sum_{\ve\beta \in \ca} \eigv_{\ve\beta} \psi_{\ve\beta}(x).\\
\end{align*}
Because the PCE basis functions $\psi_{\ve\beta}$ are of different orders, we can rewrite this into matrix form:
\begin{equation}
\frac{1}{\Nlatent - 1} \sum_{r=1}^{\Nlatent} 
\begin{pmatrix}
\tilde{\coeff}^r_{\ve\beta_1} (\tilde{\ve \coeff}^r)^T  \\
\tilde{\coeff}^r_{\ve\beta_2} (\tilde{\ve \coeff}^r)^T   \\
\vdots \\
\tilde{\coeff}^r_{\ve\beta_P} (\tilde{\ve \coeff}^r)^T 
\end{pmatrix}
\ve \eigv
= \left( \frac{1}{\Nlatent - 1} \sum_{r=1}^{\Nlatent} \tilde{\ve \coeff}^r \ (\tilde{\ve \coeff}^r)^T \right) \ve \eigv
= \left( \frac{1}{\Nlatent - 1} \ \tilde{\ve \coeff} \ \tilde{\ve \coeff}^T \right) \ve \eigv
\overset{!}{=} \lambda \ve \eigv,
\label{eq:PCE_PCA}
\end{equation}
where $\tilde{\ve \coeff}$ is the $\Nactive \times \Nlatent$-matrix of active and centered PCE trajectory coefficients. 
Defining the matrix $\ve \Sigma = \frac{1}{\Nlatent - 1} \ \tilde{\ve \coeff} \ \tilde{\ve \coeff}^T$, which is the empirical covariance matrix of the centered PCE coefficients, we see that
\cref{eq:PCE_PCA} is nothing else than principal component analysis (PCA) on the coefficients.

Note that it was necessary to apply KLE in $\ivsp$ to arrive at this equation, since the PCE basis is orthonormal in $\ivsp$ but in general not in $L^2(\cd)$.

The solution vectors $\ve \eigv^{(i)}$ to the eigenvalue problem $\ve \Sigma \ve \eigv^{(i)} = \lambda_i \ve \eigv^{(i)}$ are the PCE coefficients of the KLE eigenfunctions: $\phi_i(\ve x) = \sum_{\alp \in \ca} \eigv^{(i)}_\alp \psi_\alp(\ve x)$. 
Since the PCE basis is orthonormal, and assuming that the eigenvectors $\ve \eigv^{(i)}$ are normalized to unit norm, it follows that the eigenfunctions $\{\phi_i\}$ are orthonormal.

\subsection{Analytical computation of the realizations of the KL-RV}
\label{app:derivations_KL-RV}
Let $\lambda_i$ be an eigenvalue and 
\begin{equation}
\phi_i(x) = \sum_{\alp \in \ca} \eigv^{(i)}_\alp \psi_\alp(\ve x)
\end{equation}
the associated eigenfunction expressed in the PCE basis.
The projection of the PCE trajectories onto the KLE basis is given by
\begin{align*}
\xi_i^r 
&= \frac{1}{\sqrt{\lambda_i}} \int_\cd \tilde{\cm}^\text{PCE}_r(\ve x) \phi_i(x) f_{\ve X}(\ve x) \ \di{\ve x} 
\\
&= \frac{1}{\sqrt{\lambda_i}} \sum_{\alp \in \ca} \tilde{\coeff}^r_\alp \eigv^{(i)}_\alp 
= \frac{1}{\sqrt{\lambda_i}} (\tilde{\ve \coeff}^r)^T \ve \eigv^{(i)} \, \in \Rr	.
\end{align*}
Let 
$\tilde{\ve \coeff} \in \Rr^{\Nactive \times \Nlatent}$ the matrix of coefficients of centered PCE trajectories
and $\ve \eigv \in \Rr^{\Nactive \times \Nkl}$ the matrix of PCE coefficients of the KLE functions. 
Then we can compute the matrix $\ve\Xi \in \Rr^{\Nkl \times \Nlatent}$ of KLE coefficient realizations by
\begin{equation}
\ve\Xi =  \left(\text{diag}\left(\frac{1}{\sqrt{\lambda_1}} \enum \frac{1}{\sqrt{\lambda_\Nkl}} \right) \ve \eigv \right)^T \tilde{\ve \coeff}.
\label{eq:KL-RV_realizations}
\end{equation}

\end{document}